\documentclass[prx,twocolumn,english,superscriptaddress,floatfix,longbibliography]{revtex4-2}

\usepackage[T1]{fontenc}
\usepackage[utf8]{inputenc}
\usepackage{graphicx}
\usepackage{xcolor}
\colorlet{BLUE}{blue}
\colorlet{BLACK}{black}
\usepackage{amssymb}
\usepackage{upgreek}
\usepackage{amsmath}
\begin{document}

\title{Dissipative self-assembly of colloidal suspensions}%

\author{Jason Conradt}%
\affiliation{Department of Chemical and Biomolecular Engineering, Allan P. Colburn Laboratory, 150 Academy St., University of Delaware, Newark, Delaware 19716, USA}
\author{Eric M. Furst}%
\email[Corresponding author: ]{furst@udel.edu}
\affiliation{Department of Chemical and Biomolecular Engineering, Allan P. Colburn Laboratory, 150 Academy St., University of Delaware, Newark, Delaware 19716, USA}

\date{December 8, 2025}%

\begin{abstract}
Suspensions of paramagnetic colloids exhibit kinetic arrest in strong magnetic fields. Through a dissipative process of toggling the field on and off, suspensions self-assemble into dense and dynamic steady-state phases. Based on the domain elongation, alpha- and contour-shapes, and degree of phase separation, we construct a phase diagram using a k-means clustering analysis. We identify six characteristic structural regimes: a structureless phase, an arrested structure, sheets, ribbons, a spiky phase, and a transient fluid-fluid regime. We further report the distribution and alignment of domains and the generality of the results. We model self-assembled domain shapes using an equilibrium mean-field magnetostatic energy calculation, which predicts the surprising emergence of highly-anisotropic structures driven by the sample's confinement. 
\end{abstract}

\maketitle

\begin{figure*}[t]
\includegraphics[width=\textwidth]{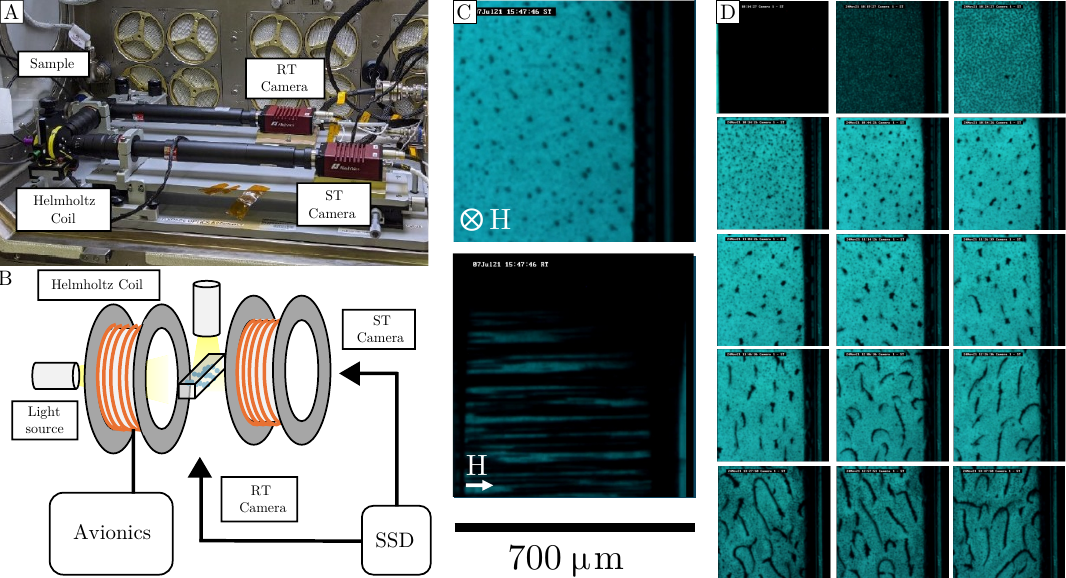}
\caption{(A) Experimental apparatus for magnetically-guided self-assembly. Equipment housed in the microgravity glovebox on the International Space Station. (B) Schematic representation of key coil assembly components. (C) Field-parallel ``ST'' and field-perpendicular ``RT'' microscopy images. The field axis, denoted $H$, corresponds to the laboratory $x$-axis. The vertical dimension on the ST view is the laboratory $z$-axis, corresponding to the length of the 50~mm capillary. The horizontal dimension on the ST view corresponds to the laboratory $y$-axis. (D) Time evolution of the suspension structure at field strength $H_0 = 2276~\mathrm{A/m}$, frequency $\nu = 2~\mathrm{Hz}$, and duty ratio $\xi = 0.20$.}
\label{fig:exp}
\end{figure*}

\section{Introduction}

Dissipative and dynamic colloidal self-assembly are part of a broad class of physical processes in which microscopic components spontaneously organize into larger structures. In colloidal systems, several distinct routes to self-assembly can be identified based on how the underlying energy landscape is defined and maintained. In \emph{passive} colloidal self-assembly, particles organize into equilibrium structures through free-energy minimization. Examples include colloidal crystals formed by entropy maximization in suspensions of hard spheres~\cite{Pusey1986,Zhu1997b} or by minimizing interaction energies between particles, including field-induced interactions \cite{yethiraj2003,Lumsdon2004,ma2013,Du2013} or polymer-mediated forces \cite{Gast1983,He2020,ofosu2024}. To overcome kinetic bottlenecks that can arrest passive assembly in gel-like or glassy states, \emph{directed} self-assembly employs external stimuli---such as electric or magnetic fields, flow, light, or chemical gradients---to reshape the energy landscape and guide structure formation~\cite{fialkowski2006,grzelczak2010b,forster2011}. In contrast, \emph{dissipative} or dynamic self-assembly describes systems that form and maintain organized structures only while energy is continuously supplied or periodically dissipated~\cite{fialkowski2006,tagliazucchi2016,arango-restrepo2019,liljestrom2019,coughlan2019}. These nonequilibrium assemblies are of particular interest because they resemble the mechanisms that sustain organization in biological systems and other forms of active matter, where structure persists through ongoing energy consumption rather than thermodynamic equilibrium \cite{sanchez2012,cates2015,bishop2023}.

Applying toggled magnetic fields to suspensions of paramagnetic colloidal particles provides a particularly tractable way to study dissipative self-assembly because particle interactions can be switched rapidly, creating a continuous and periodic input of energy \cite{martin2013,spatafora-salazar2021}. Suspensions exhibit a range of structures in toggled fields that depend strongly on its frequency and duty ratio. When the field frequency is comparable to the characteristic relaxation time of the particle structure during the off phase, toggled fields drive the formation of body-centered tetragonal (BCT) crystals \cite{swan2014}, the equilibrium structure of densely packed dipolar spheres \cite{tao1991,toor1992,Sherman2019,pal2015}. In contrast, very high frequencies (short off-times) produce networked, non-crystalline gel-like structures \cite{promislow1996,swan2012,kim2020}, while very low frequencies (long off-times) yield only transient structures that dissolve within each toggle cycle, if they form at all \cite{kim2020}. Simulations of mutually polarizing dipolar spheres reproduce these trends and additionally predict metastable coexistence regimes at lower duty ratios, where crystalline BCT cores are surrounded by a dense colloidal fluid \cite{Sherman2019}. Subsequent experiments using confocal microscopy and laser scattering revealed an additional dynamic ``wavy'' phase consisting of mobile colloidal bands with intermediate crystallinity and no equilibrium analogue \cite{kim2020}. Together, these results demonstrate that periodically driven magnetic suspensions can access a range of non-equilibrium steady-states and emergent dynamic structures that do not arise under constant magnetic fields. Importantly, recent work shows non-equilibrium steady-states in systems under toggled interactions can be described by equilibrium equations of state, revealing an unexpected connection between the two~\cite{sherman2016,sherman2019c}.

In this work, we characterize the self-assembly of paramagnetic colloidal particles subjected to toggled uniaxial magnetic fields under microgravity conditions using experiments performed on the International Space Station (ISS). Building on earlier microgravity studies that examined the roles of particle shape, volume fraction, and toggle frequency \cite{swan2012,kim2019}, we investigate how varying the toggle duty ratio---together with frequency and field strength—governs the shape and distribution of self-assembled particle aggregates. A surprisingly rich state diagram emerges. 
These experiments bridge microscopic understanding from recent theory and simulations of toggled dipolar interactions \cite{sherman2019c} with the larger length and time scales associated with aggregates approaching macroscopic dimensions. Conducting the experiments in microgravity removes the confounding effects of sedimentation, which on Earth produces quasi-two-dimensional aggregates, and instead allows direct observation of fully three-dimensional assembly.

We first outline the materials and methods, then quantify the existence of steady-state phases and the morphological characteristics of the these phases induced by changing field conditions. We characterize the emergence of distinct self-assembled aggregate shapes, distributions, and orientations of aggregates throughout the parameter space. Finally, we compare the shape dependence of the magnetostatic energy of self-assembled aggregates, which implicates the suspension confinement as an important driver of aggregate anisotropy. 
 
\section{Methods}
 
Paramagnetic colloidal suspensions were prepared by suspending polystyrene spheres containing magnetite nanoparticles (MyOne Dynabeads) in ultra-pure water (resistivity 18.2~M$\Omega\cdot$cm). Three separate suspensions were prepared, each corresponding to one of three nominal particle diameters: $1.05\,\upmu\mathrm{m}$, $0.48\,\upmu\mathrm{m}$, or $0.26\,\upmu\mathrm{m}$. The beads have a magnetic susceptibility of $\chi = 1.4$ and a saturation magnetization of $M_{\mathrm{sat}} = 40\,\text{kA}\,\text{m}^{-1}$. Each suspension sample had a volume fraction of $\phi = 5 \times 10^{-3}$ (0.5\%).

Each sample used approximately $20\,\upmu\mathrm{L}$ of suspension loaded into a rectangular glass capillary (Wale Apparatus, part no.\ 8270-50; width 0.70~mm, wall thickness 0.14~mm, length $\approx 50$~mm). One end of the capillary was sealed with epoxy (Hardman Epoxy \#04003), and the other end was flame-sealed using an acetylene torch. The sealed capillary was then placed inside a sample vial enclosure at the center of a Helmholtz coil assembly in the Microgravity Science Glovebox, see Figure~\ref{fig:exp}. Near the midpoint between the coils, the magnetic field was uniform along the central coil axis. Two cameras were positioned to view the microstructure near the capillary center: an ``RT'' camera oriented perpendicular to the field direction, and an ``ST'' camera oriented parallel to it (Figure~\ref{fig:exp}A). 

The InSPACE-4 campaign comprised 96 flight test points spanning multiple particle types, field strengths, frequencies, and duty ratios \cite{PSI-177}. A concise summary of the experimental matrix is given in the Supplemental Material. 
\begin{figure*}[t]
\centering
\includegraphics[width=5in]{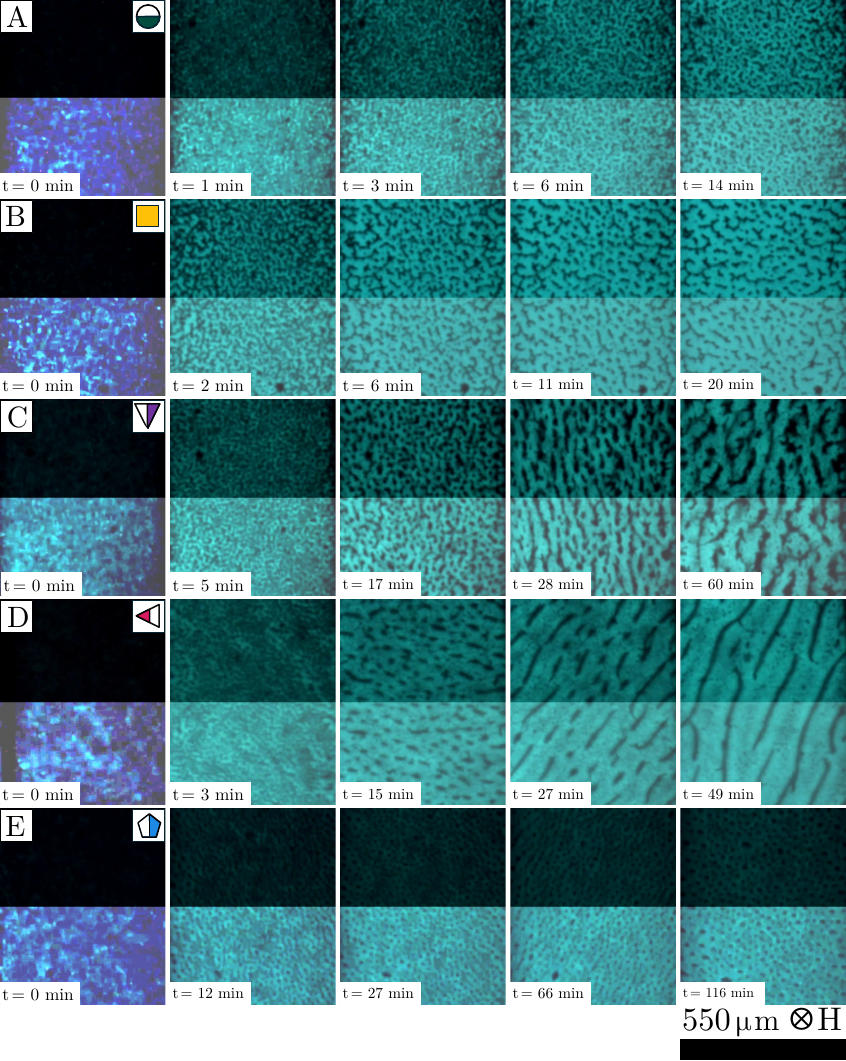}
\caption{(A--E) Sample kinetics of structure formation for various suspension steady states. Top halves show raw images, bottom halves are brightness- and contrast-equalized to highlight structure. The dark regions are two-dimensional projections of self-assembled colloidal aggregations, the bright regions are particle-dilute background. Steady states are differentiated by field conditions which are encoded into the symbols on the leftmost panel of each row; the legend for these symbols is given in Fig.~\ref{fig:kinetics}.}
\label{fig:kinetics2}
\end{figure*}
In ground-based experiments, sedimentation forces cause particles and aggregates to accumulate rapidly on the walls of their container. This is difficult to mitigate because the paramagnetic particles are denser, and the particles used in these experiments are almost twice as dense as their carrier fluid, water (specific gravity $s.g.=1.8$). The onset and timescale of sedimentation can be estimated with a gravitational P\`eclet number~\cite{whitmer2011}
\begin{equation}
\mathrm{Pe}_g = \frac{\tau_{D}}{\tau_g} = \frac{4\pi\,\Delta\rho\,R_a^4\,g_{\text{eff}}}{4\,k_B T},
\end{equation}
where $\tau_g$ is the time for a particle to move one diameter under acceleration $g_{\text{eff}}$, $\Delta\rho$ is the density difference between the particles and the fluid, and $R_a$ is the effective radius of a particle aggregate. Because $\mathrm{Pe}_g \propto R_a^4$, even small density mismatches will drive $Pe_{g}\gg 1$. We show in the Supplemental Material that ambient and acute accelerations on the ISS are not expected to influence the distribution of particle structures in our experiments.

The coil assembly produces magnetic fields at the center of the sample with magnitudes $H_0$ ranging from $627$ to $2276~\mathrm{A/m}$. The field is expected to vary by less than $1~\mathrm{A/m}$ through the capillary along the $x$-axis, corresponding to the ST view. Along the $z$-axis, corresponding to the vertical direction of the ST view, the field is similarly uniform within the camera's field of view. However, this is expected to decay to zero near the ends of the 50~mm capillary, resulting in field gradients along the full capillary length.
Details about this calculation are given in the  Supplemental Material.
The applied magnetic field is perturbed by Earth's geomagnetic field. We provide detailed calculations regarding the effect of geomagnetism on the ISS in the  Supplemental Material . We show that the peak disturbances induced by Earth's geomagnetic field along the $\mathbf{H}_0$ axis are $\mathcal{O}(10) ~\mathrm{A/m}$, and oscillate with complex patterns of irregular periodicity on the order of 30 to 90 minutes. Thus, the influence of the geomagnetic field is small, on the order of 1\% of the applied field, and transient over the course of experiments, which lasted 50 minutes or more. 

\begin{figure*}[t]
\centering
\includegraphics[width=6in]{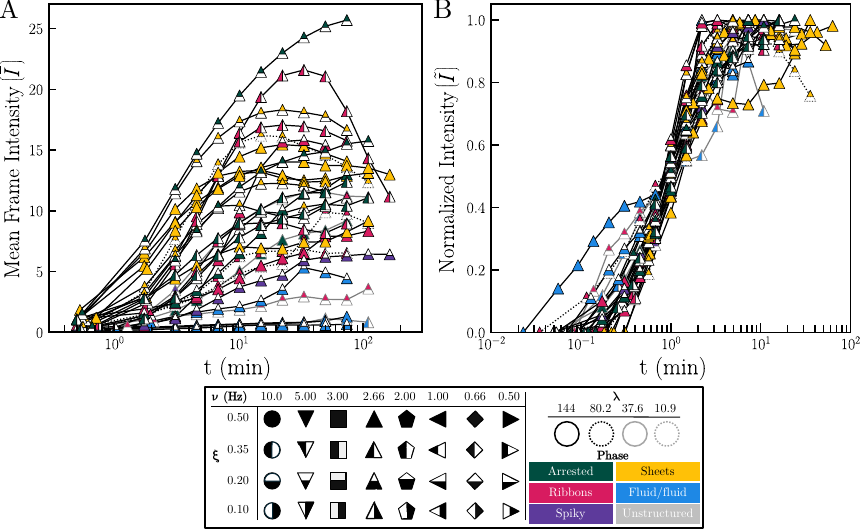}
\caption{(A) Growth kinetics of the suspension are shown with the average transmitted light intensity through the sample cell ($\bar{I}$) versus experiment time. Average intensities are computed over one-minute intervals. (B) The collapsed curves where intensity is normalized and bound to $[0,1]$ through transformation to $\tilde{I}$ where $\tilde{I}=(\bar{I}-\bar{I}_\mathrm{min})/\max(\bar{I}-\bar{I}_\mathrm{min})$. Experiment time is made dimensionless as $t/t_{1/2}$ where $t_{1/2}$ is the time at which $\tilde{I}$ reaches half of its maximum.}
\label{fig:kinetics}
\end{figure*}

Paramagnetic particles magnetize linearly with the external field at these field strengths. We define a dimensionless field strength relative to thermal energy using the relative dipole ratio~\cite{spatafora-salazar2021,kim2020,swan2012,swan2014}
\begin{equation}
\lambda = \frac{\pi \mu_0\,R^3\,\chi\,H_0^2}{9\,k_B T},
\end{equation}
where $k_B$ is the Boltzmann constant, $T$ is the absolute temperature, $\mu_0$ is the vacuum permeability, $\chi$ is the particle's magnetic susceptibility, $H_0$ is the applied field magnitude, and $R$ is the particle radius. When $\lambda \gg 1$, thermal (Brownian) energy is much less than the induced magnetic interactions between particles, and particles rapidly form aggregates.
The coil assembly produces square waves of the magnetic field, with a frequency range $0.25\leq \nu \leq 20$~Hz and a duty ratio range $0.10\leq \xi \leq 0.5$. The duty ratio is defined as the fraction of each toggle period during which the field is on $t_{\text{on}}$,
\begin{equation}
\xi = \frac{t_{\text{on}}}{t_{\text{on}} + t_{\text{off}}},
\end{equation}
We also express the off-time, $t_{\text{off}}$, in terms of $\xi$ and the toggling frequency $\nu$ (in Hz) by
\begin{equation}
t_{\text{off}} = \frac{1 - \xi}{\nu}.
\end{equation}
By adjusting $t_{\text{off}}$, we control the duration in which Brownian motion can act between field pulses~\cite{swan2014}. We compare this against the diffusive timescale~\cite{swan2014,kim2020}
\begin{equation}
\tau_D = \frac{6\pi \eta\,R^3}{k_B T},
\end{equation}
where $\eta$ is the fluid viscosity. 

The ratio $t_{\text{off}}/\tau_D$ indicates how far the particles can diffuse when the field is off \cite{swan2014,kim2020}. At $t_{\text{off}}/\tau_D=1$, an isolated particle will on average diffuse the length of its own radius. Thus, choosing $\xi$, $\nu$, and $H_0$, we tune the suspension energy scale and diffusive timescales in a way that is relatable to different external field strengths and particle sizes. For particle diameters ranging from $0.26~\upmu\mathrm{m}$ to $1.05~\upmu\mathrm{m}$, $t_{\text{off}}/\tau_D$ spans roughly 0.038 to 360. 

\section{Structure evolution to a steady-state}
\label{sec:transmission}
\begin{figure*}[t]
\includegraphics[width=\textwidth]{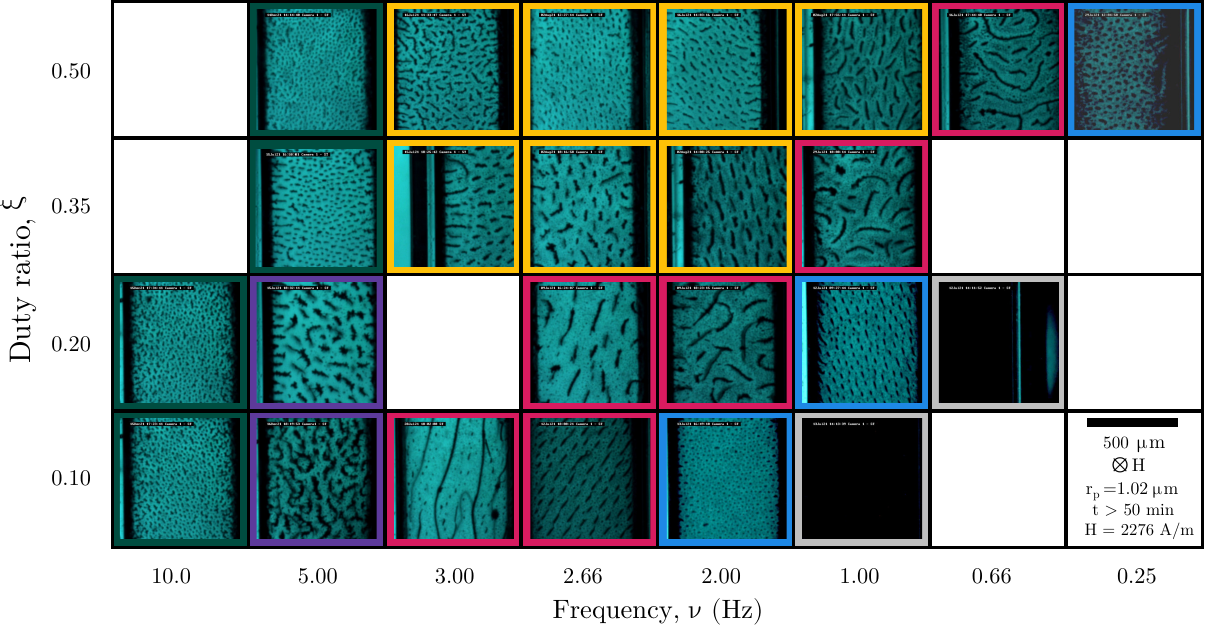}
\caption{Video micrographs of the steady-state suspension structure. Toggle frequency ($\nu$) and duty ratio ($\xi$) are varied at a constant field strength $H_0 = 2276~\mathrm{A/m}$. Red, purple, orange, green, blue, and gray backgrounds denote the structure as being ribbons, spiky, sheets, arrested, fluid-fluid, and structureless phases, respectively, as identified in Section \ref{sec:phasediagram}. The dimensions of each image are $700~\upmu\mathrm{m}\times 700~\upmu\mathrm{m}$. The particle diameter is $2R = 1.05~\upmu\mathrm{m}$.}
\label{fig:steadystates}
\end{figure*}

For field strengths of at least $1164~\mathrm{A/m}$ and at square-wave toggling frequencies exceeding $0.25~\mathrm{Hz}$, initially disordered suspensions aggregate into clusters. In all experiments where structure emerged, aggregation became detectable within ten minutes ($\sim 1000\tau_D$) of applying the field. Representative time series for several field protocols are shown in Fig.~\ref{fig:kinetics2}A--E (field conditions are encoded by the symbols at left; see the legend in Fig.~\ref{fig:kinetics}). The micrographs are taken in the field-parallel ST view (Fig.~\ref{fig:exp}C), so they show two-dimensional projections of structures that extend along the field axis. Aggregates appear as dark regions, while the surrounding dilute fluid transmits green light. Structure does not resolve in the field-perpendicular RT view because field-aligned aggregates elongate toward the capillary depth and occlude the illumination.  (The marker color here and in subsequent plots gives the steady state phase assigned to that structure, an assignment we discuss in detail in Sec.~\ref{sec:phasediagram}.)

Continued toggling drives the suspension toward steady-state morphologies that depend on the toggle protocol. All structured runs pass through a similar early morphology within $\sim 10$ minutes (comparable to the steady state in Fig.~\ref{fig:kinetics2}A and visible in the second column of Fig.~\ref{fig:kinetics2}B--E) which likely reflects an interconnected, labyrinthine demixing pattern \cite{swan2012,swan2014}. Under some conditions, aggregates continue to evolve after this initial stage. The dominant change is in aspect ratio; domains flatten perpendicular to the field while their number density decreases, consistent with continued aggregation over the next 10 to 20 minutes. This flattening is apparent between the second and fourth columns for several runs in Fig.~\ref{fig:kinetics2}B--D. At longer times, the suspension approaches one of several characterstics: domains become thin and long, or they thicken and develop pronounced surface texture. These slower transformations typically take on the order of one hour and are accompanied by more dynamic aggregate motion. A distinct growth mode (Fig.~\ref{fig:kinetics2}E) exhibits slow, continuous coarsening over the full experiment duration, producing diffuse columnar or ellipsoidal domains with low contrast; the persistent haze suggests that many free particles and small aggregates remain dispersed in the continuous phase.

We are primarily interested in the steady-state structures in toggled fields. While the kinetics of self-assembling colloids are of independent interest, and can constrain the mechanisms that drive aggregation \cite{vicsek1984}, here we use kinetic measurements mainly to verify that each experiment reaches a steady state. A simple and robust proxy for aggregate formation is the mean transmitted light through the sample over the course of the experiment. 
The transmitted intensity is sensitive not only to the extent of aggregation but also to continued morphological evolution. For example, progressive flattening can increase the projected dark area in the image plane (and, if aggregate volume is approximately conserved, would be accompanied by changes along the field axis that are not resolved in the ST projection). Alternatively, a growing population of small or unresolved aggregates could attenuate transmitted light without producing strong contrast associated with the resolved domains. For these reasons, we use the stability of image intensity as a practical indicator that the time-averaged suspension structure has reached a steady state.

Stability of the mean transmitted intensity is evaluated by measuring the systematic drift in intensity in time. For each run we compute the frame-averaged transmitted intensity, $\bar{I}(t)$, in one-minute bins and then quantify residual evolution by comparing successive five-minute averages; by the end of the experiment, the mean percent change between adjacent five-minute intervals typically falls below $1\%$. Figure~\ref{fig:kinetics}A shows $\bar{I}(t)$ for all experiments, while Figure~\ref{fig:kinetics}B replots the same data using the normalized intensity
\begin{equation}
\tilde{I} = \frac{\bar{I} - \bar{I}_{\mathrm{min}}}{\max(\bar{I} - \bar{I}_{\mathrm{min}})},
\end{equation}
and the rescaled time $t/t_{1/2}$, where $t_{1/2}$ is the time at which $\tilde{I}$ reaches half of its maximum value. Across the parameter space, $\bar{I}(t)$ typically plateaus or becomes bounded and weakly oscillatory on times of order $50$--$250$~min, corresponding to $5000\tau_D$ to $25000\tau_D$ (see  Supplemental Material ).

For the representative $700~\upmu\mathrm{m}\times700~\upmu\mathrm{m}\times700~\upmu\mathrm{m}$ observation volume at $\phi=5\times10^{-3}$, approximately $5\times10^6$ particle are expected to populate the the field of view, while the largest aggregates discussed below contain order $10^5$ particles. The structures and experiment times studied here are larger than can presently be treated by direct mutually polarizing Brownian dynamics simulations~\cite{Sherman2019,sherman2019c}.

\section{Steady-state structures}
\label{sec:structure}
We connect structure and dynamics to different field protocols by measuring the suspension meso- and macrostructure using quantitative image analysis \cite{conradt2025}. By \emph{mesostructure} we mean the geometry of individual aggregates---their lengths, thicknesses, and surface textures---while \emph{macrostructure} refers to their spatial arrangement and mutual orientation across the field of view. 
To perform the analysis, we take micrographs from the steady states of each experiment by selecting the brightest frame by mean pixel intensity within one toggle period of the desired time, corresponding to the end of the on-phase of a toggle cycle. This ensures consistent images across the data set.

We begin this section by quantifying the degree of anisotropy induced in the aggregate shapes by defining a major and minor dimension for aggregates and measuring these parameters for all aggregates in each experiment. We follow with a similar analysis of aggregate surface shape, defined with an $\alpha$-shape method. We conclude with a characterization of the macrostructure with a two-dimensional Voronoi method to measure the separations of complex shapes and an equivalent ellipsoid approach to determine the ensemble aggregate alignment across experiments. The ultimate goal of this analysis is to identify the distinct steady-state morphological phases from which we can draw a quantitatively grounded phase diagram.

\subsection{Aggregate anisotropy}

Inspection of steady-state micrographs suggests that the clearest descriptor of mesostructure lies in the anisotropy of aggregates. Aggregates tend to be thin, and much of what separates each phase lies in the relative dimensions of the self-assembled aggregates imaged along the field axis (the ST view). We extract lengths and widths by analyzing the morphological skeletons and contours of objects imaged at steady state. This is performed by first converting the RGB-color space micrograph to a binary image followed by extracting each morphological point set. We then calculate the major dimension of each object, its length, by finding the longest contiguous chain in its contour. Its minor dimension, the thickness, is computed by calculating the mean minimum distances between each point on the skeleton and the contour. By grounding the calculation in these point sets, we get an accurate measurement of the major and minor dimension of an object that is not affected by the object's tortuosity or anisotropy. This is a significant improvement over more common approaches such as the calculation of image moments or hand measurement. A detailed review of the methods are presented in Ref.~\cite{conradt2025}.

\begin{figure*}[t]
\centering
\includegraphics[width=\textwidth]{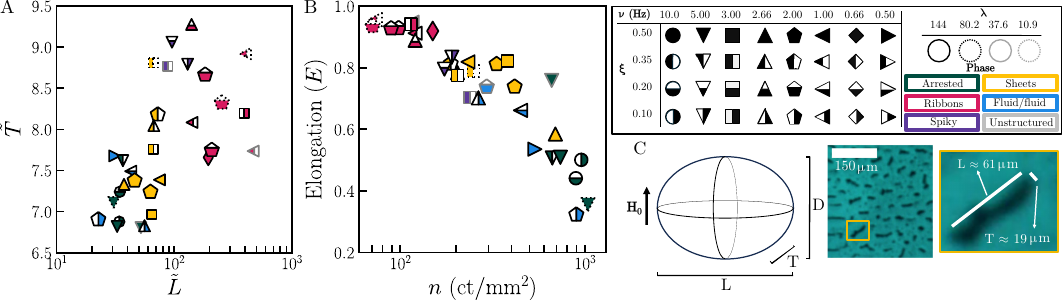}
\caption{(A) Area-weighted average dimensionless lengths ($\langle \tilde{L}\rangle$) and thicknesses ($\langle \tilde{T}\rangle$) of aggregates imaged with the field-parallel (ST) camera. These dimensions are nondimensionalized by the particle diameter, $2R=1.05~\upmu\mathrm{m}$. (B) Area-weighted average elongation, $\langle E\rangle$, as a function of aggregate number density in the micrograph. As $\langle E\rangle$ increases, the number of objects decreases, implying continued aggregation. (C) Schematic of the triaxial ellipsoid used to approximate aggregate geometry. The semi-axis $a$ always runs along the field axis, while $b$ and $c$ are the in-plane semi-axes; the measured in-plane thickness, $T$, is less than or equal to the measured length, $L$. The particle radius is given by $R$. }
\label{fig:lenswidths}
\end{figure*}

Figure~\ref{fig:lenswidths}A shows the area-weighted average dimensions for all experiments, which we cast as those of a representative ellipsoid per Figure~\ref{fig:lenswidths}C. For an individual aggregate $i$, we denote the in-plane major and minor dimensions by $L_i$ and $T_i$, respectively, and define its elongation as
\begin{equation}
E_i=1-\frac{T_i}{L_i}.
\end{equation}
The plotted quantities are the area-weighted means $\langle L\rangle$, $\langle T\rangle$, and $\langle E\rangle$, with $\langle \tilde{L}\rangle=\langle L\rangle/(2R)$ and $\langle \tilde{T}\rangle=\langle T\rangle/(2R)$. The mean aggregate thickness $\langle \tilde{T}\rangle$ is nearly constant, varying by no more than about 20\% between any two experiments, and ranges from about 6.5 to 9.5. The range of thicknesses is significant, as it is near to the critical capture radius derived by the accumulation of Brownian forces on a particle over $\tau_D$, which describes the distance at which the thermal force scale matches the magnetic interaction force scale. This is given by~\cite{swan2012} 
\begin{equation}
R_c=2^{\frac{7}{6}}R\lambda^{\frac{1}{3}},
\end{equation} 
and for our data using one micron particles, we find $R_c/(2R)$ ranges from about 5.1 to 12. Thus the observed aggregate thicknesses lie within the same order of magnitude as $R_c/(2R)$. This observation extends prior observations about the significance of the critical capture radius in describing suspension behavior. In similar microgravity experiments, Swan and co-workers found a transition in the aggregation kinetics as aggregates grew to the thickness of the capture radius, moving from diffusion-limited coarsening to a regime characterized resembling the late-stage kinetics of spinodal decomposition~\cite{swan2012}. 

Unlike $\langle \tilde{T}\rangle$, the area-weighted mean major dimension $\langle \tilde{L}\rangle$ changes significantly, ranging from $20 \lesssim \langle \tilde{L}\rangle \lesssim 600$. Figure~\ref{fig:lenswidths}B shows that $\langle E\rangle$ increases systematically as the number density of aggregates decreases. This relationship is not obvious. The suspension could in principle have maintained similar number densities while aggregates became thin and elongated. That flattening occurs alongside continued aggregation and fewer total clusters in the suspension implies a connection between aggregate volume and shape. We explore this connection in more detail in Sec.~\ref{sec:energy}.

\subsection{Aggregate surface texture}

\begin{figure}
\centering
\includegraphics[width=0.9\columnwidth]{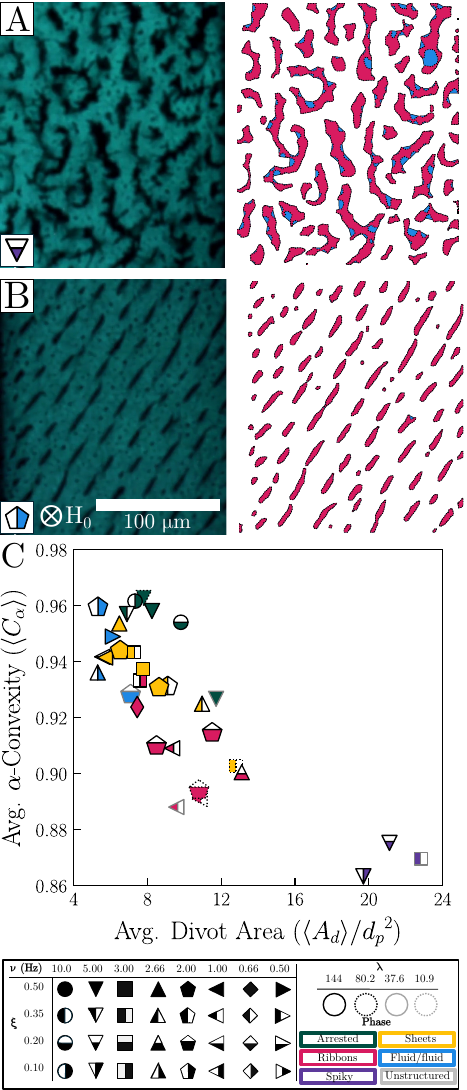}
\caption{Sample analyses and statistics of aggregate interfacial texture versus applied field conditions. (A) Sample experiment corresponding to the minimum $C_\alpha$ (labelled B on (A)) rendered as a set of $\alpha$-shapes and their underlying contours. (B) Sample experiment corresponding to the maximum $C_\alpha$ (labelled C on (A)) rendered as a set of $\alpha$-shapes and their underlying contours. (C) Average $\alpha$-convexities of aggregates in suspensions as a function of effective field strength and characteristic diffusion time. Background colors correspond to approximate areas where each phase is observed, corresponding to the legend in Figure~\ref{fig:lenswidths}A. }
\label{fig:roughness}
\end{figure}

Length and thickness alone do not capture how corrugated the aggregate boundaries are; some phases exhibit nearly smooth outlines, whereas others develop highly textured, ``spiky'' surfaces. Figures~\ref{fig:roughness}A and~\ref{fig:roughness}B illustrate suspension states whose aggregates are rough (A) and smooth (B), where rough specifically refers to the bends and points on the surface of the aggregate whose lateral extent is on the order of, or smaller than, the aggregate thickness. We quantify roughness by generating the \emph{alpha shapes} of the aggregate cross-sections and computing from them metrics describing surface texture. 

An $\alpha$-shape is a generalization of the convex hull and represents the exterior hull of an $\alpha$-complex, a construct closely related to the Delaunay triangulation of a point set~\cite{edelsbrunner1994,gardiner2018,asaeedi2017,borgoni2021}. In essence, we take the point set that describes each object's contour and draw connections between any two points which intersect with a circle of radius $\alpha^{-1}$ and do not contain any additional points. In the limit of an infinitely large circle, this returns the convex hull of the point set. By making $\alpha$ progressively larger (thus decreasing the connecting circle diameter) we progressively reduce the convex hull to the contour points themselves. If we choose $\alpha$ judiciously, we have a tool by which we can distinguish surface texture from the bends and twists of anisotropic objects. A detailed discussion with more explanatory figures is provided in Ref.~\cite{conradt2025}.

Using each aggregate's minor dimension, its thickness $T_i$, as the value for $\alpha^{-1}$, we assert that connections spanning a breadth greater than the thickness are macroscopic undulations in the body rather than bumps or ridges. Each shape in Figures~\ref{fig:roughness}A and~\ref{fig:roughness}B shows a binary object segmented from the base image (colored in red, outlined in solid black) alongside the associated $\alpha$-shape (outlined in dotted black). The space between the $\alpha$-shape and contour is colored blue. It is clear that there are many more blue regions in the objects processed in~\ref{fig:roughness}B than we see in~\ref{fig:roughness}C. We can use this as a distinguishing characteristic that allows us to quantify the relative roughness of two objects. We characterize surface roughness first by straightforwardly tabulating the average areas of the divots (the blue regions in~\ref{fig:roughness}A/B) for each aggregate, $A_d$, averaged across all aggregates in the image. We then introduce a new parameter, the $\alpha$-convexity $C_\alpha$, which is the ratio of the contour area ($A_{0}$) to the $\alpha$-shape area ($A_\alpha$),
{\setlength{\abovedisplayskip}{6pt}
 \setlength{\belowdisplayskip}{6pt}
 \setlength{\abovedisplayshortskip}{6pt}
 \setlength{\belowdisplayshortskip}{6pt}
 \begin{equation}
 C_\alpha = \frac{A_0}{A_\alpha}.
 \end{equation}
}
\noindent 
The $C_\alpha$ parameter is analogous to a shape's convexity, the ratio of its contour area to convex hull area, which is frequently used in image characterization problems~\cite{blott2008}. The divot areas express the typical size of surface perturbations while the $\alpha$-convexities show the relative magnitudes of these ridges and valleys to the overall aggregate size.  

Figure~\ref{fig:roughness}C shows the $C_\alpha$ values against $\langle A_d \rangle$ for all experiments, where markers are colored per the phase diagram discussed in Sec.~\ref{sec:phasediagram}. The $\alpha$-convexities range from approximately 0.86 to 0.96, and divot areas normalized by the square of the particle diameter ($2R = 1.05~\upmu\mathrm{m}$) range from approximately 4 to 24. Changes in $C_\alpha$ could come from increasing the size and frequency of divots in the surface contour at constant length and thickness, creating more or less circular bodies which will weight the area ratio of the alpha shape to base contour towards one, or some combination thereof. The clear inverse relationship between $\langle A_d \rangle/(2R)^2$ and $\langle C_\alpha \rangle$ shows that the increasing roughness implied by decreasing $\langle C_\alpha \rangle$ is largely caused by the formation of larger surface undulations. These surface perturbations are subdued at higher frequencies, field strengths, and duty ratios. 

\section{Suspension macrostructure}
\label{sec:macrostructure}

Next, we quantify the suspension macrostructure. The primary feature of interest is how aggregates are distributed in the sample capillary. We expect that aggregates should repel laterally as they do not appear to overlap with one another, suggesting aggregates are aligned next to each other along the field axis. 
Magnetostatic dipolar interactions between aggregates scale with their volumes and the inverse cube of their separation, $U_{ij}\sim V_{a,i}V_{a,j}/r_{ij}^3$~\cite{nye1984,jackson2003}. Thus, coalescing into smaller numbers of larger aggregates may be energetically favorable if it also increases the characteristic separation between them. Because the interaction decays more rapidly with distance than it grows with aggregate volume, separation may be a significant aspect of suspension structure. 

\subsection{Aggregate distribution}

\begin{figure}[h]
\includegraphics[width=2.75in]{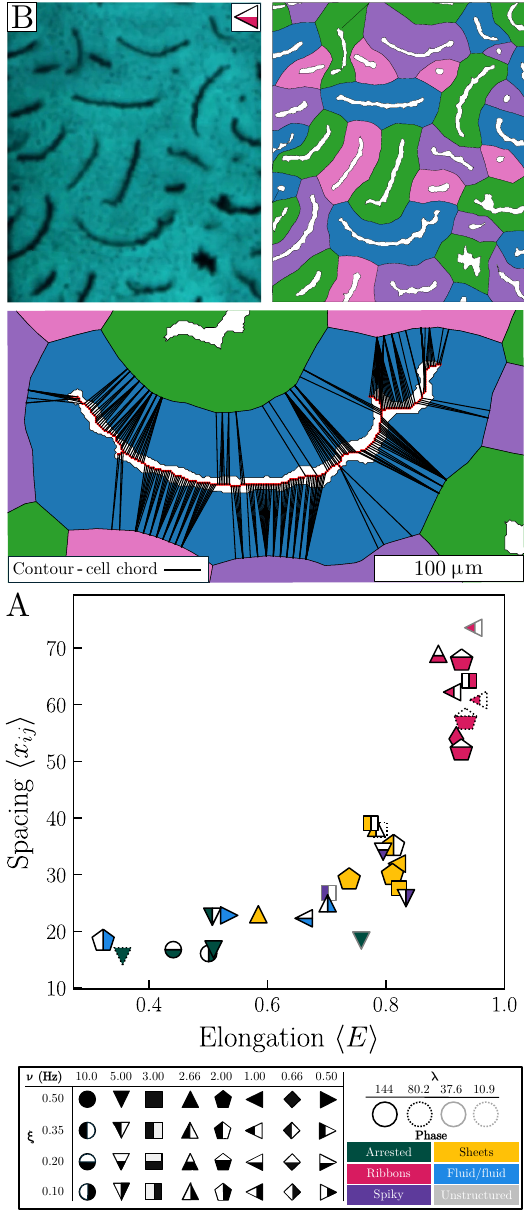}
\caption{Sample analyses and statistics of aggregate distribution versus applied field conditions. (A) Sample area Voronoi analysis and separation estimate. The micrograph shown in B is discretized into a set of regions differentiated by color using their contour points. The space owned by a single aggregate is taken as the average contour-cell chord length between an aggregate's surface and the border of its Voronoi cell. (B) Average inter-aggregate spacing computed as the mean chord distance, $\langle x_{ij} \rangle$, between the aggregate's morphological skeleton and the Voronoi cell produced by the two-dimensional morphological contour of all aggregates. }
\label{fig:spacing}
\end{figure}

Quantifying inter-aggregate separation is practically challenging for anisotropic objects because their distances vary as a function of position on the cluster and their relative orientations. The average separation between two sheet- and ribbon-like shapes is often poorly approximated by the separation of their centroids. We address this challenge by using the morphological contour of aggregates in each image to create an area-based Voronoi diagram that accounts for the two-dimensional extent of each object. A detailed explanation and analysis of the method is given in Ref.~\cite{conradt2025}. Figure~\ref{fig:spacing}A shows an example of such an analysis, with the average of the chords drawn in the zoomed-in frame being used to deduce the space ``belonging'' to each object and twice the average chord length corresponding to inter-aggregate separations. 

Figure~\ref{fig:spacing}B shows the distribution of separations as a function of elongation. At the lowest elongations, aggregates are distributed uniformly, centered at approximately 20 particle diameters apart. The more elongated aggregates extend this mean separation to about 25 to 40 particle diameters. Structures which exhibit the lowest aggregate number densities and greatest elongation have the highest inter-aggregate spacings, ranging from $50(2R)$ to $75(2R)$. 
The consolidation of  aggregates has the effect of increasing the average space between particle clusters and decreasing the aggregate number density. 
The observed trend---fewer, larger aggregates that are farther apart---is therefore consistent with the simple scaling argument based on magnetostatic repulsion presented above. The steady state structures are not characterized by the clustering of large aggregates, but rather maximize their separation. This also suggests that magnetostatic interaction energies diminish as a fraction of total suspension energy as aggregates elongate.

\subsection{Aggregate alignment}

\begin{figure}
\centering
\includegraphics[width=3in]{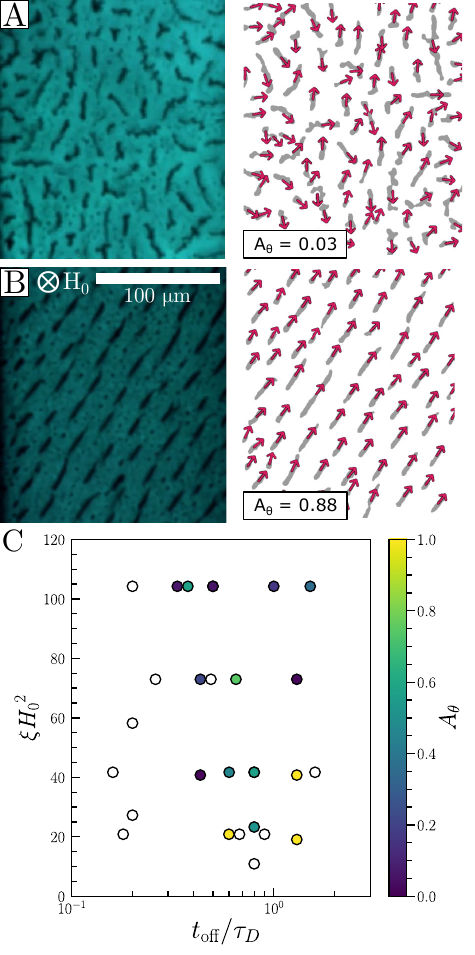}
\caption{Sample analyses and statistics of aggregate alignment versus applied field conditions. (A) Sample experiment of highly unaligned aggregates. (B) Sample experiment with similar length and thickness, but exhibiting high alignment. Brightness has been enhanced by 50\% for clarity. (C) Alignment factors versus effective energy scale ($\xi {H_0}^2$) and dimensionless off-time ($t_{\mathrm{off}}/\tau_D$) with phases colored per the legend given in Figure~\ref{fig:lenswidths}A. }
\label{fig:alignment}
\end{figure}

A final macrostructural parameter examined is that of relative aggregate orientations. We notice in some experiments that aggregates align in the field-perpendicular plane (the $y$--$z$ image plane), as though ordered by a magnetic field component normal to the applied field axis. For example, the micrograph shown in Figure~\ref{fig:alignment}A illustrates elongated aggregates with no clear, mutual alignment, while the micrograph in Figure~\ref{fig:alignment}B shows qualitatively similar aggregates that are highly aligned. To quantify this behavior, we compute for each anisotropic body the orientation of an ellipse that best represents the object's shape. We use a moments-based approach common to the computer vision literature~\cite{gonzalez2018} and, in parallel, a directly parameterized ellipse fit using a least-squares method~\cite{fitzgibbon1999}. An overlap criterion selects which ellipse's orientation is used to estimate that of the underlying body, as detailed elsewhere~\cite{conradt2025}. We quantify relative orientations with an alignment factor $A_\theta$
\begin{equation}
A_\theta=\frac{45^\circ-\langle \theta_{ij}\rangle}{45^\circ},
\end{equation}
where $\theta_{ij}$ is the pairwise relative orientation wrapped to $[0^\circ,90^\circ]$, and $45^\circ$ is the mean pairwise angle for a random ensemble of ellipses. A large $A_\theta$ implies small relative angles and therefore high alignment; $A_\theta=1$ corresponds to perfect mutual alignment, $A_\theta=0$ to a random ensemble, and negative values indicate preferential mutual misalignment.

Figure~\ref{fig:alignment}C shows experiments plotted on a phase diagram whose points are color-scaled to $A_\theta$. Several experiments are absent from this diagram because their aggregates were too circular to estimate orientations reliably. Orientations contribute to image statistics only if objects have aspect ratio greater than $1.5$ and cross-sectional area greater than $144\,(2R)^2$, and we accept frames only if fewer than 30\% of objects are rejected by these filters. Under these criteria, eleven experiments are disqualified (white circles on Figure~\ref{fig:alignment}C) and fifteen are viable. 

Among the viable runs, six exhibit $A_\theta\lesssim 0.25$ (near-random ordering), four show $A_\theta\gtrsim 0.50$ (marked mutual alignment), and the remaining five fall in between. The alignment is unlikely to originate from chance because a random distribution has $\langle \theta_{ij}\rangle=45^\circ$ with a standard deviation of approximately $12^\circ$, thus values $A_\theta>0.5$ represent several standard deviations of shift relative to the random case.

We considered several hypotheses to explain alignment among anisotropic aggregates. One possibility is that alignment reflects a lower-energy arrangement that maps directly onto steady-state phase. However, we find no simple phase-by-phase correspondence in this dataset. Regions of higher alignment do appear more often at higher off-times and lower effective fields, but the trend is not consistent enough to support a strong conclusion. 

We also considered whether experiment time influences the alignment. If mutual alignment were simply the endpoint of slow rearrangement for elongated aggregates, one might expect $A_\theta$ to increase with run duration. We do not observe that behavior, even when mesostructures remain approximately steady for multiple hours. 

We also consider whether alignment may be influenced by geomagnetic perturbations in the field-perpendicular plane. 
We estimate the in-plane geomagnetic magnitude $|\mathbf{h}_{yz}|=\sqrt{H_y^2+H_z^2}$ for each run using the International Geomagnetic Reference Field model over the course of each experiment. Measuring the orientations of aggregates compared with the geomagnetic field orientation over time, we then average $|\mathbf{h}_{yz}|$ over the analysis window for each run and compare it with the internal alignment $A_\theta$ (mutual alignment between aggregates) and the field-alignment factor $A_{g}$ (alignment of aggregates to the in-plane field direction measured in the image)
\begin{equation}
A_g = \frac{45^\circ - \langle \theta_{ig} \rangle}{45^\circ},
\label{eq:orderparam}
\end{equation}
where 
\begin{equation}
\theta_{ig} = \left| \theta_{i} - \theta_{g} \right|,
\label{eq:orientation}
\end{equation}
and $\theta_{i}$ is the orientation of an object $i$ and $\theta_{g}$ is the direction of the geomagnetic field. Like $A_\theta$, $A_g$ equals 1 for perfect alignment with the in-plane geomagnetic field direction, 0 for random orientations, and negative values indicate preferential misalignment.

\begin{figure*}[t]
\centering
\includegraphics[width=6.75in]{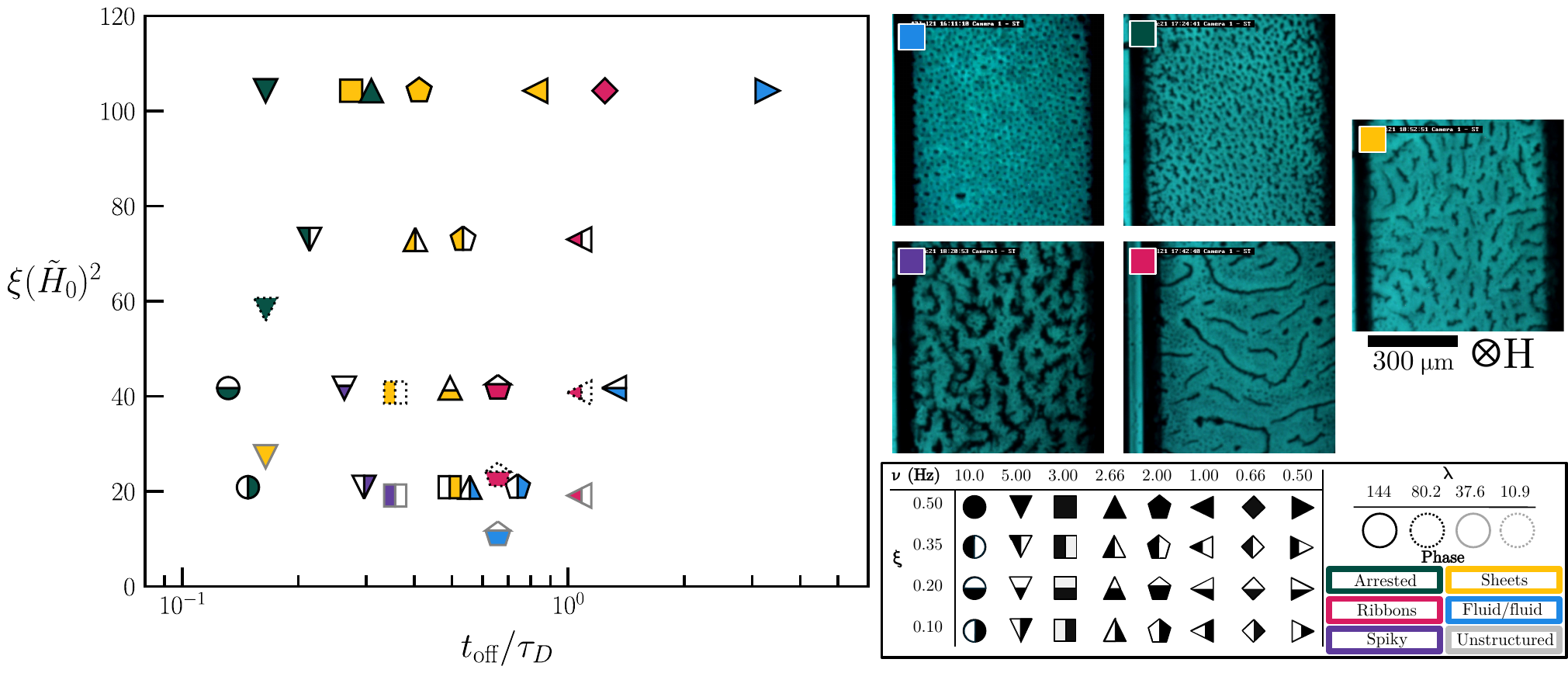}
\caption{Phase diagram for the steady-state suspension structures of paramagnetic suspensions in toggled magnetic fields under microgravity. Structures are mapped onto the diagram along the vertical axis in terms of the energy scale of the applied field written as $\xi {\tilde{H}_0}^2$ where the duty ratio is given as $\xi=t_\mathrm{on}/(t_\mathrm{on}+t_\mathrm{off})$ and $\tilde{H}_0$ is a normalized effective field given by $\tilde{H}_0=\sqrt{R^3  \mu_0 / (k_B T)}\,H_0$. Along the horizontal axis, experiments are catalogued by their dimensionless field-off time $t_\mathrm{off}/\tau_{D}$.}
\label{fig:phasediagram}
\end{figure*}

Pearson tests determine whether stronger in-plane geomagnetic fields correspond to stronger alignment with that field, as well as whether a stronger in-plane field corresponds to stronger mutual alignment between aggregates. We find that $A_{g}$ increases with $\langle|\mathbf{h}_{yz}|\rangle$ ($r=0.493$, two-sided $p=0.0048$), implying that stronger in-plane geomagnetic fields are associated with aggregates pointing along the in-plane field direction. However, this effect by itself is too small to explain the differences in alignment between experiments.  $A_\theta$ does not correlate with $\langle|\mathbf{h}_{yz}|\rangle$ ($r=-0.084$, $p=0.653$). Finally, we find $A_\theta$ does not covary with $A_{g}$ ($r=0.050$, $p=0.788$). 
Ultimately, we do not yet have a concrete explanation for the global, mutual alignment that sometimes occurs in these suspensions. High $A_\theta$ appears in some experiments with very long aggregates and in some less structured phases at long off-times, suggesting that morphology and dynamics may both matter. The geomagnetic field may bias the orientation of individual aggregates through a weak torque, while strong mutual alignment likely depends on suspension dynamics and residence times. In some aggregation regimes, aggregates may simply not persist long enough in one location to achieve strong mutual alignment.

\section{Morphological phase diagram}
\label{sec:phasediagram}

In this Section, we combine the descriptors presented in Sections \ref{sec:transmission} and \ref{sec:structure} to construct a state diagram. We categorize the suspension steady states into phases defined by the average texture and anisotropy of mesostructures, weighted by aggregate cross-sectional area. We use the normalized average divot area, $\langle A_d \rangle/(2R)^2$, as a metric for texture, the average elongation, $\langle E \rangle$, as a metric of anisotropy, and the mean transmitted light intensity, $\langle I \rangle$, as a broad metric for the consolidation of the suspension. The range of each metric is rescaled to $[0,1]$ across all runs, and the resulting feature vectors are organized into $k$ discrete groups using the $k$-means clustering algorithm. Structureless suspensions are excluded from the clustering because no aggregate shapes emerge from which we can extract the relevant metrics.

We varied the number of clusters in the range $3 \le k \le 8$ and, for each set of clusters, computed the Xie--Beni (XB) index, a cluster-validity measure that compares the total within-cluster variance to the squared separation between the closest pair of cluster centroids. This explicitly penalizes the creation of diffuse or redundant clusters with nearly identical centroids~\cite{ikotun2025,xie1991}. This metric is well suited to small data sets with overlapping clusters~\cite{ikotun2025}. A detailed overview of the clustering procedure and evaluation is given in the  Supplemental Material . 

In Figure~\ref{fig:phasediagram}, we show the resulting phase diagram expressed in terms of a dissipative timescale and an effective magnetostatic energy scale, $t_{\mathrm{off}}/\tau_D$ and $\xi \tilde{H}_0^{\,2}$, respectively, where $t_{\mathrm{off}}$ is the field off-time, $\tau_D$ is the single-particle self-diffusion time at infinite dilution, and $\tilde{H}_0=\sqrt{R^{3}\mu_0/(k_B T)}\,H_0$ is a dimensionless field strength. Because $\lambda\propto H_0^2$ for fixed particle properties, the vertical axis $\xi\tilde{H}_0^{\,2}$ is equivalent to a duty-ratio-weighted dipolar energy scale while making the role of periodic driving explicit. Marker color is defined by the phase assigned by the $k$-means algorithm. Phases occupy largely contiguous regions of the $(\xi \tilde{H}_0^{\,2},\,t_{\mathrm{off}}/\tau_D)$ plane, indicating that these axes capture the dominant physics governing structure. Consistent with prior toggled-interaction theory and simulation, scaling the field strength with duty ratio collapses experiments performed at multiple dipolar coupling strengths~\cite{sherman2019c}. The smooth progression of morphologies further suggests that aggregate mesostructures can be tuned deterministically by adjusting the dissipative timescale and effective energy scale. 

The morphologically defined phases are visually distinct. The suspension structures are titled as the arrested phase, sheets phase, ribbons phase, spiky phase, and fluid--fluid phase; examples of each are provided in Figure~\ref{fig:phasediagram} and are defined by the legend. The phases are named primarily to reflect the aggregate geometries they yield, though some of the titles reflect inferences about the underlying processes leading to those structures. 

In the remainder of this Section, we provide more details for each phase, beginning with the arrested and fluid-fluid phases that form in the limits of short and long off-times and then turning to the intermediate regime where sheets, ribbons, and spiky aggregates emerge. 

\subsection{Short and long off-times}
  
\begin{figure}[t]
\centering
\includegraphics[width=\columnwidth]{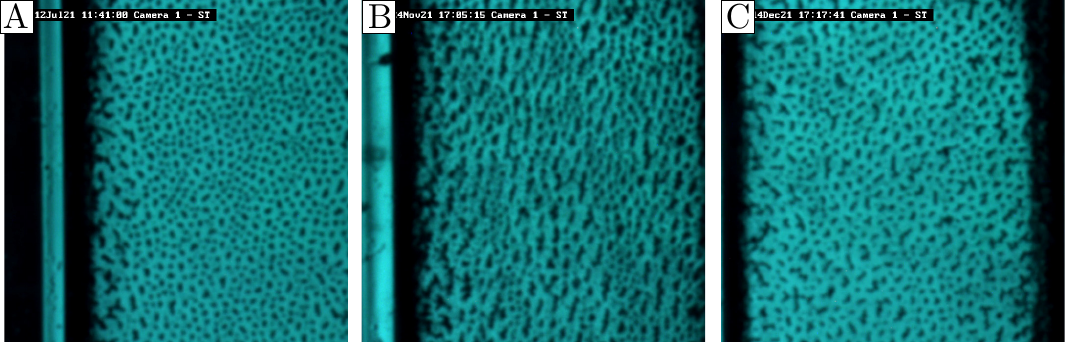}
\caption{Representative steady-state field-parallel (ST) micrographs of the arrested (short-$t_{\mathrm{off}}/\tau_D$) regime in Fig.~\ref{fig:phasediagram}. The applied field is directed normal to the image plane and the image scale is $(700\times704$ $\mu$m; dark regions are particle-rich aggregates and bright regions are the dilute continuous phase. (A) $H_0=2276~\mathrm{A/m}$, $\nu=10.0~\mathrm{Hz}$, $\xi=0.20$ ($\xi\tilde{H}_0^{\,2}=45.8$, $t_{\mathrm{off}}/\tau_D=0.121$). (B) $H_0=2276~\mathrm{A/m}$, $\nu=10.0~\mathrm{Hz}$, $\xi=0.10$ ($\xi\tilde{H}_0^{\,2}=22.9$, $t_{\mathrm{off}}/\tau_D=0.136$). (C) $H_0=1701~\mathrm{A/m}$, $\nu=5.00~\mathrm{Hz}$, $\xi=0.50$ ($\xi\tilde{H}_0^{\,2}=63.9$, $t_{\mathrm{off}}/\tau_D=0.151$).}
\label{fig:arrested}
\end{figure}

At low $t_\mathrm{off}/\tau_D \lesssim 0.2$, the suspension develops into a distribution of many small aggregates with relatively low cross-sectional aspect ratios, typically on the order of one to two, which correspond to mean elongations ranging from approximately 0.3 to 0.6. These experiments appear on the left side of Figure~\ref{fig:phasediagram}, denoted by green markers. This phase emerges rapidly, achieving a steady state within ten minutes of field onset; see, for example, the structure evolution shown in Figure~\ref{fig:kinetics}A. Three examples of an arrested-phase steady state are shown in Figure~\ref{fig:arrested}. The onset of this suspension structure occurred where $t_\mathrm{off}/\tau_D\rightarrow0.1$, with all observations occurring between $t_\mathrm{off}/\tau_D=0.1$ and $0.3$.

An analogous phase has been observed both during a previous set of microgravity self-assembly experiments, where a ``system-spanning'' structure was noted by \citet{swan2012}, and in gravity, where related ``percolated'' structures have been identified~\cite{promislow1997c,kim2020}. Our images are consistent with a similar morphology: the dot-like structures observed at high toggle frequencies may be cross-sectional projections of percolated columns, while the static hazy regions between columns are suggestive of unresolved branches and out-of-plane connections. However, the lack of an RT view in these experiments prevents a direct confirmation of such connectivity. The kinetic sequence is likewise consistent with arrest at an early stage through which other structured phases pass, similar to late-stage spinodal-like morphologies discussed previously~\cite{semwal2022,rendos2022,kim2020,bauer2015,bauer2016,swan2014,swan2012,sherman2018}. 

A plausible explanation for the locked, immobile structure that emerges at low $t_\mathrm{off}$ is that the off-phase is too brief for substantial restructuring within a toggle cycle. The toggling process has been described as an oscillating thermostat for a paramagnetic phase that uses cycles of melting ($t_\mathrm{off}$) and freezing ($t_{\mathrm{on}}$) to overcome kinetic barriers created by strong, anisotropic interactions~\cite{swan2012,spatafora-salazar2021}. At sufficiently small $t_\mathrm{off}$, surface particles may not diffuse far enough from their original positions to adopt new configurations upon reapplication of the field. On the suspension scale, this is consistent with a gel-like set of networked columnar aggregates. 

We can also make limited inferences about the particle-scale structure by drawing analogies to experiments performed in standard gravity and to representative simulations. Small-angle light scattering on sedimented paramagnetic suspensions under comparable toggled-field conditions revealed diffuse, six-fold scattering peaks in the percolated state, consistent with short-range, chain-like local crystallinity but little long-range crystalline order~\cite{kim2020}. Together with the system-spanning, percolated morphology and arrested coarsening, this is suggestive of a kinetically arrested, gel-like network. Simulations that include both mutual polarization and hydrodynamic interactions produce analogous percolated gel-like phases under constant fields~\cite{Sherman2019}; theoretical analyses of dissipative systems with oscillatory potentials show that, in the high-frequency limit, toggled interactions are equivalent to a constant, time-averaged potential~\cite{risbud2015,tagliazucchi2016}. These comparisons suggest that the structures we produce in high-frequency fields are analogous to those produced in constant or time-averaged fields. 
\begin{figure}[t]
\centering
\includegraphics[width=\columnwidth]{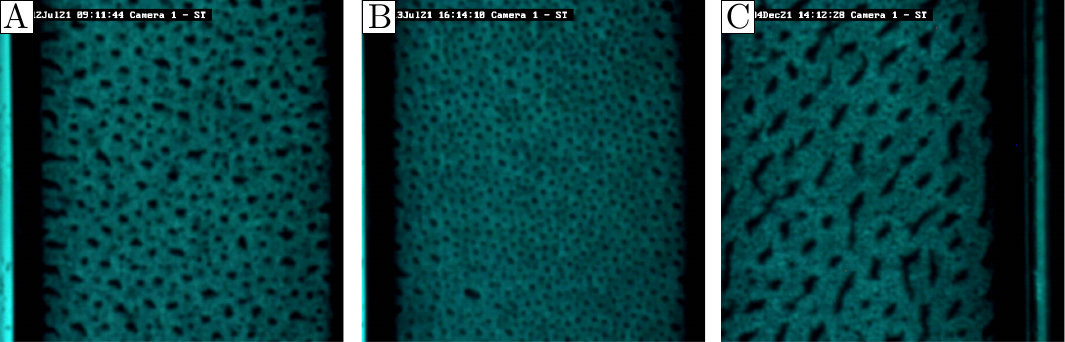}
\caption{Representative steady-state field-parallel (ST) micrographs of the fluid--fluid regime at large $t_{\mathrm{off}}/\tau_D$ (blue region in Fig.~\ref{fig:phasediagram}), characterized by diffuse, continuously evolving particle-rich regions and a persistent background haze. The applied field is directed normal to the image plane and the image scale is $(700\times704$ $\mu$m; dark regions are particle-rich and bright regions are dilute fluid. (A) $H_0=2276~\mathrm{A/m}$, $\nu=2.00~\mathrm{Hz}$, $\xi=0.20$ ($\xi\tilde{H}_0^{\,2}=45.8$, $t_{\mathrm{off}}/\tau_D=0.603$). (B) $H_0=2276~\mathrm{A/m}$, $\nu=2.00~\mathrm{Hz}$, $\xi=0.10$ ($\xi\tilde{H}_0^{\,2}=22.9$, $t_{\mathrm{off}}/\tau_D=0.679$). (C) $H_0=1164~\mathrm{A/m}$, $\nu=2.00~\mathrm{Hz}$, $\xi=0.20$ ($\xi\tilde{H}_0^{\,2}=12.0$, $t_{\mathrm{off}}/\tau_D=0.603$).}
\label{fig:fluid}
\end{figure}

At the opposite extreme of large $t_\mathrm{off}/\tau_D \gtrsim 2$, the off-phase is sufficiently long that aggregates melt significantly, losing most of their spatial correlations between field pulses. Under these conditions, we observe a slowly coarsening fluid--fluid regime where aggregate surfaces remain in flux through the entire on- and off-phases of the toggle cycle. Here, less light is transmitted through the sample, illustrated in Figure~\ref{fig:kinetics}G and Figure~\ref{fig:fluid}, reflecting a higher fraction of unassociated particles over an entire toggle period. In contrast to the arrested gel, the structure in this regime never becomes rigid, so we regard it as a transitory steady state maintained by the repeated melting and reformation of aggregates during each cycle. The aggregates themselves are perhaps better considered centers of high particle concentration, as it isn't clear that densely packed cores of particles actually form.

The threshold off-times for this suspension structure increases as the time-averaged field strength decreases, with this state being noted at $t_\mathrm{off}/\tau_D\approx0.6$ for $\xi(\tilde{H}_0)^2\approx20$, increasing to $t_\mathrm{off}/\tau_D\approx1.5$ for $\xi(\tilde{H}_0)^2\approx40$, then finally being noted at $t_\mathrm{off}/\tau_D\approx4$ for $\xi(\tilde{H}_0)^2\approx100$; though parameters were not sampled thoroughly around the last point and this may not reflect the phase boundary well. This trend reflects the increased ability for stronger fields to "catch" particles during the on-phase through increased $R_c$. 

\subsection{Intermediate off-times}

In an intermediate window of $t_\mathrm{off}/\tau_D$, aggregate surface particles are expected to be able to explore new configurations during the off-phase without completely losing the memory of the aggregate as a whole. We believe this description applies to the sheets, ribbons, and spiky morphologies highlighted in Figure~\ref{fig:phasediagram}.
 
\subsubsection{Sheets phase}
 
\begin{figure}[t]
\centering
\includegraphics[width=\columnwidth]{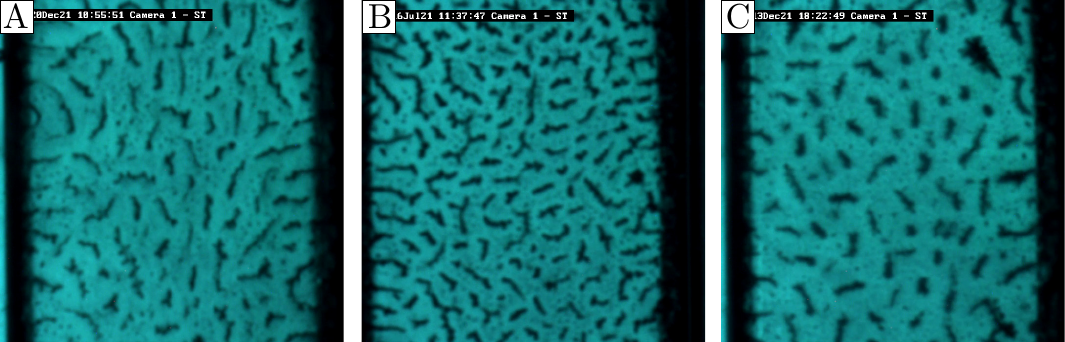}
\caption{Representative steady-state field-parallel (ST) micrographs of the sheets phase (orange region in Figure~\ref{fig:phasediagram}), showing stable, flattened and elongated sheet-like aggregates. The applied field is directed normal to the image plane and the image scale is $(700\times704$ $\mu$m; dark regions are particle-rich aggregates and bright regions are the dilute continuous phase. (A) $H_0=2276~\mathrm{A/m}$, $\nu=2.00~\mathrm{Hz}$, $\xi=0.50$ ($\xi\tilde{H}_0^{\,2}=114$, $t_{\mathrm{off}}/\tau_D=0.377$). (B) $H_0=2276~\mathrm{A/m}$, $\nu=3.00~\mathrm{Hz}$, $\xi=0.50$ ($\xi\tilde{H}_0^{\,2}=114$, $t_{\mathrm{off}}/\tau_D=0.251$). (C) $H_0=1701~\mathrm{A/m}$, $\nu=3.00~\mathrm{Hz}$, $\xi=0.35$ ($\xi\tilde{H}_0^{\,2}=44.8$, $t_{\mathrm{off}}/\tau_D=0.327$).}
\label{fig:sheets}
\end{figure}

At $t_\mathrm{off}/\tau_D\approx0.2$, there exists a boundary between the arrested phase and a phase characterized by stable, flat, elongated structures which resemble sheets of particles. Figure~\ref{fig:lenswidths}A shows that this phase exhibits intermediate aspect ratios, and Figure~\ref{fig:lenswidths}B shows a decrease in number density associated with this morphological change. The decrease in number shows that these structures result from continued aggregation beyond the initial, percolated structure. The sheets phase is additionally distinct from the arrested phase in that it lacks the hazy regions between domains that suggest a branched, percolated network. Figure~\ref{fig:sheets} shows sample images of this phase at steady state.

Breakup of system-spanning structures into layers of sheet-like aggregates has been observed in prior microgravity experiments~\cite{swan2014} and in sedimented suspensions of paramagnetic colloids~\cite{kim2020,promislow1996}. This was thought to occur because of a balancing of interfacial magnetic energies in the dense (and possibly crystalline) particle aggregate, contributions from its shape-dependent demagnetizing field, and gravitational loading of sedimented layers, which favored flattened structures. Interestingly, our results show that the sedimentation effect has little to do with the onset of the anisotropic flattened structures, as the sheets phase occurs under nearly identical toggle conditions as a similar anisotropic phase observed under a gravitational field~\cite{kim2020}.

\subsubsection{Ribbons phase}
 
\begin{figure}[t]
\centering
\includegraphics[width=\columnwidth]{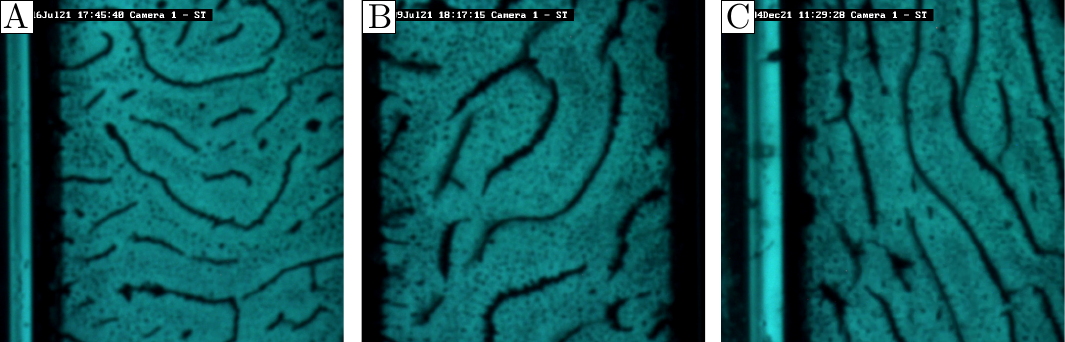}
\caption{Representative steady-state field-parallel (ST) micrographs of the ribbons phase (red region in Fig.~\ref{fig:phasediagram}), comprising highly elongated, winding ribbon-like domains. The applied field is directed normal to the image plane and the image scale is $(700\times704$ $\mu$m); dark regions are particle-rich aggregates and bright regions are the dilute continuous phase. (A) $H_0=2276~\mathrm{A/m}$, $\nu=0.66~\mathrm{Hz}$, $\xi=0.50$ ($\xi\tilde{H}_0^{\,2}=114$, $t_{\mathrm{off}}/\tau_D=1.14$). (B) $H_0=2276~\mathrm{A/m}$, $\nu=2.00~\mathrm{Hz}$, $\xi=0.20$ ($\xi\tilde{H}_0^{\,2}=45.8$, $t_{\mathrm{off}}/\tau_D=0.603$). (C) $H_0=1701~\mathrm{A/m}$, $\nu=2.00~\mathrm{Hz}$, $\xi=0.20$ ($\xi\tilde{H}_0^{\,2}=25.6$, $t_{\mathrm{off}}/\tau_D=0.603$).}
\label{fig:ribbons}
\end{figure}

Between the fluid--fluid and sheets phases on Figure~\ref{fig:phasediagram}, a dynamic phase of elongated domains forms. We refer to these elongated, winding structures as ``ribbons'' in recognition of both their long contours and irregular bends. Three micrographs of the ribbon phase produced under various field conditions are shown in Fig.~\ref{fig:ribbons}. These aggregates continue to grow in length until they acquire distinctive dynamical behaviors including splitting, merging, and the accretion of small particle clusters. As captured by the data in Figure~\ref{fig:lenswidths}A, aggregates in this phase are significantly longer than those in the sheets phase, and some ribbons can extend the entire length of the sample cell. Analogous structures of colloidal bands or "zig-zags" have been observed in thin, sedimented layers of self-assembled paramagnetic suspensions~\cite{guell1988,kim2020,kach2024}. The data presented here constitute the first observation of this type of structure under microgravity.

\subsubsection{Spiky phase}

\begin{figure}[t]
\centering
\includegraphics[width=\columnwidth]{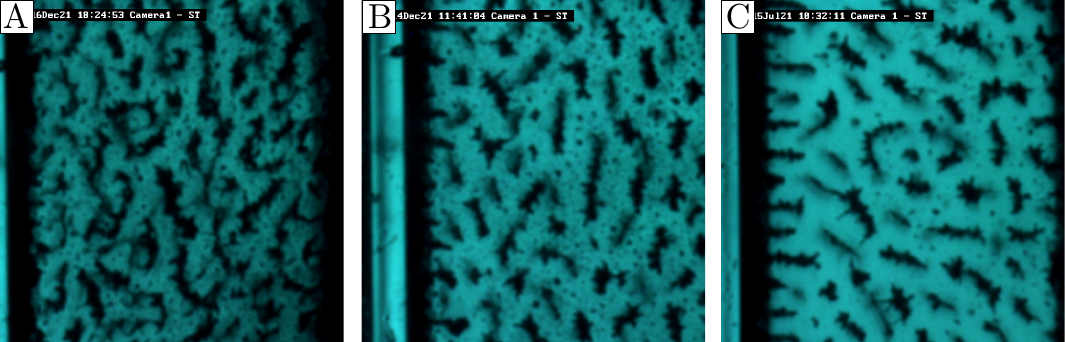}
\caption{Representative steady-state field-parallel (ST) micrographs of the spiky phase (purple region in Fig.~\ref{fig:phasediagram}), distinguished by strongly textured aggregate boundaries with pronounced protrusions/undulations. The applied field is directed normal to the image plane and the image scale is $(700\times704$ $\mu$m; dark regions are particle-rich aggregates and bright regions are the dilute continuous phase. (A) $H_0=2276~\mathrm{A/m}$, $\nu=5.00~\mathrm{Hz}$, $\xi=0.10$ ($\xi\tilde{H}_0^{\,2}=22.9$, $t_{\mathrm{off}}/\tau_D=0.272$). (B) $H_0=1164~\mathrm{A/m}$, $\nu=3.00~\mathrm{Hz}$, $\xi=0.35$ ($\xi\tilde{H}_0^{\,2}=21.0$, $t_{\mathrm{off}}/\tau_D=0.327$). (C) $H_0=2276~\mathrm{A/m}$, $\nu=5.00~\mathrm{Hz}$, $\xi=0.20$ ($\xi\tilde{H}_0^{\,2}=45.8$, $t_{\mathrm{off}}/\tau_D=0.241$).}
\label{fig:spiky}
\end{figure}

Building on the idea that our morphologically defined phases also correspond to microstructural differences, we can comment on the origin of the highly textured ``spiky'' phase that is observed at the intersection of the arrested, sheets, and ribbons phases. Examples of these suspension structures are shown in Fig.~\ref{fig:spiky}. Stable undulations in the aggregate contour imply a balance of particle fluxes everywhere on the aggregate surface within a toggle cycle. The migration of particles around the aggregate surface will be affected by multiple energetic and kinetic variables, such as the local dipolar interaction energy and demagnetizing field, the curvature-dependent interfacial free energy at the aggregate--fluid interface, and the surface diffusion coefficient governing lateral motion of particles along the interface. For example, protrusions will have a bias towards growth because they possess larger local capture regions; this effect competes against surface diffusion in the off-phase which will tend to flatten the surface. Simultaneously, capillary forces stemming from magnetic surface energies may drive the formation of bump interfaces, analogous to capillarity-driven growth instabilities at colloidal crystal--fluid interfaces in repulsive systems~\cite{gast1991}. Any stable surface structure will represent a non-equilibrium steady state with respect to the transport limitations and oscillatory, energetic driving force which lead to the destructuring and restructuring of the aggregate surface.

This picture is consistent with prior work in toggled fields where coarsening proceeds largely by surface diffusion, and with the observation that columns aligned with the field undergo a Rayleigh--Plateau instability in which magnetic surface energies drive undulations that are resisted by a frequency-dependent viscosity set by collective diffusion~\cite{bauer2015}. In this frame, aggregates are like a dense fluid phase whose particle mobility is tuned by $t_{\text{off}}/\tau_D$ and temperature by $\sqrt{\xi}\,\tilde{H}_0$. In those cases, these instabilities lead to the breakup of system-spanning aggregates into shorter, stable aggregates. The reassembled surface during the on-phase will be affected by anisotropic, dipolar surface energies~\cite{toor1992} that may preferentially direct surfaces to form along specific crystallographic directions; the disassembling surface in the off-phase will have to overcome anisotropic, hard sphere surface energies~\cite{marr1994}. These nuances are further supplemented by the complex hydrodynamic interactions governing collective diffusion in the off-phase, which will be a function of the exact interfacial structure. 

We draw an analogy to a simpler but related case, the process of nanowire spheroidization. This is a process by which columnar aggregates or crystals of metallic nanoparticles can either decompose into discrete spherical aggregates or arrest into more complex geometries~\cite{nichols1976,gorshkov2019}. Stable structures are selected by the system temperature, and it therefore does not need to incorporate dissipative timescales. It is a balance between surface energies and thermally driven surface diffusion which samples alternative surface configurations~\cite{nichols1976}. Breakup and restructuring are capillary effects analogous to a Rayleigh--Plateau instability. Notably, anisotropic surface energies and mobilities lead to preferred neck shapes that are set by low-energy facets; these also alter the mobility of particles at these regions~\cite{gorshkov2019}. This leads to cases of metastable lobed structures not predicted by isotropic theory~\cite{gorshkov2019}. Surface energy anisotropy leads to metastable undulated surface structures. Similar growth instabilities have been reported in colloidal crystals, where anisotropic surface kinetics generate faceted protrusions and branched morphologies~\cite{gast1991}.

We hypothesize that the stability of bumpy surfaces noted for sheets, ribbons, and especially the spiky phases has a similar physical origin. In our analogy, we tune a correlate to effective temperature when we adjust $\sqrt{\xi}\,\tilde{H}_0$ and particle mobility with $t_{\text{off}}/\tau_D$. The diffusion of the off-phase of the toggled waveform is important for sampling surface structures, guaranteeing particles remain local to an aggregate, and possibly biasing the surface structure with induced buckling-like instabilities. However, it is anisotropy in the surface energy that renders some surface orientations stable, leading to undulated aggregate textures. We hypothesize that the spiky phase represents a non-equilibrium structure that enhances interfacial undulations without pinch-offs along the aggregate body or frequent ejection of small aggregates.

\begin{figure*}[t]
\centering
\includegraphics[width=6.5in]{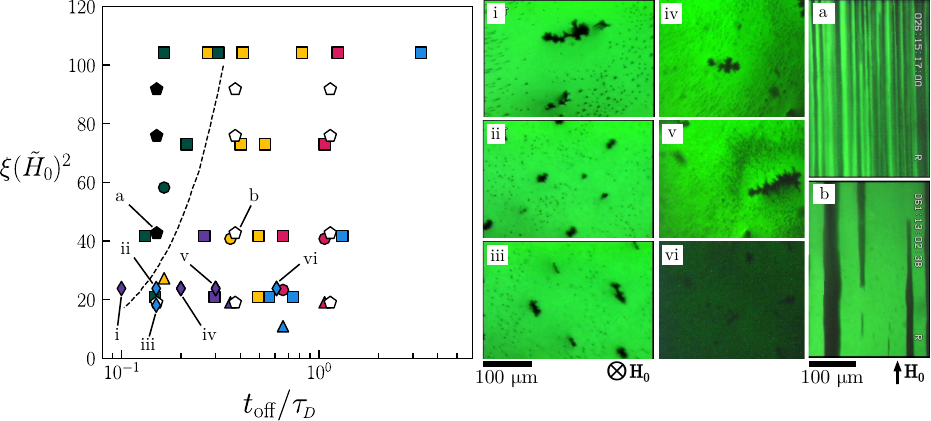}
\caption{InSPACE-1 experiments (diamond symbols) with $2R=0.66~\upmu\mathrm{m}$, $\phi=0.2\%$ and InSPACE-2 experiments with $2R=1.05~\upmu\mathrm{m}$ and $\phi=0.65\%$ (pentagon symbols; closed symbols indicate percolation and open symbols phase separation as reported in \cite{swan2012}) are overlaid on the phase diagram reported in Fig.~\ref{fig:phasediagram}. All experiments shown here are performed at duty ratio $\xi=0.50$. Images i--vi are from InSPACE-1, and images a and b are from InSPACE-2. 
The labeled experiments are (i) $H=2082~\mathrm{A/m}$ ($\lambda=37.6$), $\nu=30~\mathrm{Hz}$.
(ii) $H=2082~\mathrm{A/m}$ ($\lambda=37.6$), $\nu=20~\mathrm{Hz}$.
(iii) $H=1816~\mathrm{A/m}$ ($\lambda=28.4$), $\nu=20~\mathrm{Hz}$.
(iv) $H=2082~\mathrm{A/m}$ ($\lambda=37.6$), $\nu=15~\mathrm{Hz}$.
(v) $H=2082~\mathrm{A/m}$ ($\lambda=37.6$), $\nu=10~\mathrm{Hz}$. 
(vi) $H=2082~\mathrm{A/m}$ ($\lambda=37.6$), $\nu=5~\mathrm{Hz}$.
(a) $H=1500~\mathrm{A/m}$ ($\lambda=51.5$), $\nu=5~\mathrm{Hz}$.
(b) $H=1500~\mathrm{A/m}$ ($\lambda=51.5$), $\nu=2~\mathrm{Hz}$.
}
\label{fig:phasediagramIS12}
\end{figure*}

\section{Generality of the phase diagram}

Our study builds on two earlier ISS campaigns, InSPACE-1 and InSPACE-2. InSPACE-2 used the same magnetic particles used in this work (InSPACE-4) with multiple volume fractions of $\phi = 0.40\%$, $0.48\%$, $0.56\%$, and $0.65\%$ under toggled magnetic fields at frequencies $\nu = 0.66$ -- $30~\mathrm{Hz}$, amplitudes $H_0 = 1000$ -- $2000~\mathrm{A/m}$, and a duty ratio $\xi = 0.50$. Unlike the square cross-section of the capillaries in InSPACE-1 and 4, InSPACE-2 used rectangular capillaries (2~mm wide along $x$, 0.2~mm deep along $y$, with a length of 50~mm), enabling high-resolution field-perpendicular (RT) imaging. The data were collected in 2008 and 2009 and are archived in NASA’s Physical Sciences Informatics (PSI) database~\cite{PSI-74}. 
 
InSPACE-1 used square capillaries (1~mm $\times$ 1~mm $\times$ 50~mm), conferring similar confinement to our InSPACE-4 data set. Instead of polystyrene beads, these experiments used ferrofluid droplet emulsions (superparamagnetic iron-oxide cores in SDS/octane droplets dispersed in D$_2$O, particle susceptibility $\chi\approx1.5$) with particle diameters 0.31, 0.40, and $0.66~\upmu\mathrm{m}$. Toggling frequencies spanned 0.66 to 30~Hz at a constant duty ratio $\xi=0.50$, and most imaging was in the field-parallel ST view. Notably, fields in this study were discontinuously applied; the project scientists would manually switch the field between steady and toggled during the experiment. Two ISS campaigns were flown (2003 and 2006; the 2003 runs are archived on NASA PSI~\cite{PSI-76}).
 
These datasets complement ours by resolving structures with the field-orthogonal RT view (InSPACE-2) and by providing insights about the effect of constant fields for comparison (InSPACE-1). They also allow us to verify our phase diagram under different experiment conditions. The axes, $t_{\mathrm{off}}/\tau_D$ and $\xi\tilde{H}_0^{\,2}$, capture particle size and susceptibility, but not $\phi$ or confinement. The variation in volume fractions should affect the ground state of our suspensions as transitions among equilibrium crystal/fluid coexistences occur only at high $\phi$, and under our conditions the equilibrium is fluid/BCT coexistence (or simple fluid at low fields)~\cite{sherman2018,hynninen2005}. However, the mesostructures expected may be affected by $\phi$; for example experiments under gravity show that if $\phi$ is too low, assembly may not progress beyond isolated chains~\cite{fermigier1992}. 

Differences in capillary geometry are expected to influence both the structures we can resolve and whether system-spanning columns remain stable or break up. When the capillary depth along the field direction exceeds a critical axial wavelength, a Rayleigh--Plateau-type instability is expected to fragment system-spanning columns into multiple layers of shorter columns. Because the InSPACE-1 and InSPACE-2 capillaries are longer along the field axis, this instability is permitted in those experiments~\cite{bauer2016} and was directly observed in InSPACE-2~\cite{swan2012}.

We first take the percolation threshold reported in Fig.\ 3A of Ref.~\cite{swan2012}, which separated capillary-spanning structures (closed pentagon symbols) from collapsed or phase separated structures (open pentagon symbols), and map it onto our phase diagram. This boundary is shown on our diagram as the dotted black line in Fig.~\ref{fig:phasediagramIS12}. RT-view examples of capillary-spanning and non-spanning structures are shown in Figs.~\ref{fig:phasediagramIS12}a and~\ref{fig:phasediagramIS12}b. This curve does an excellent job separating the arrested phase region from sheets/spiky phases of the InSPACE-4 experiments. This confirms that the transition from a system-spanning structure to isolated sheets (captured from the lateral RT view in InSPACE-2) is reflective of the transition we see from our arrested structure to sheets and spiky phases from the field-parallel ST view in InSPACE-4.
 
Next, using the same morphological classifiers described above (combining $\langle E\rangle$, $\langle A_d\rangle/(2R)^2$, and $\langle I\rangle$), we map InSPACE-1 onto the $(t_{\mathrm{off}}/\tau_D,\ \xi\tilde{H}_0^{\,2})$ diagram (Fig.~\ref{fig:phasediagramIS12}). The InSPACE-1 results showing anisotropic structures that were difficult to interpret at the time, map reasonably well onto our phase diagram. Structures in Fig.~\ref{fig:phasediagramIS12}i, iv, and v were labeled as spiky, whereas ii, iii, and vi were labeled fluid--fluid. For panels i, v, and vi these assignments are qualitatively reasonable, while the remaining panels do not resemble the phase definitions established for InSPACE-4. This discrepancy underscores the sensitivity of our morphological classifier to experimental context. The low particle volume fraction yields fewer aggregates per unit cross-section; combined with the camera’s smaller field of view, each frame contains too few aggregates to support robust statistics. Lower $\phi$ also increases the mean transmitted intensity because a larger fraction of the image is bright, dilute fluid.
 
\section{Time-averaged magnetostatic aggregate energy}
\label{sec:energy}

We next seek to understand why aggregates tend to self-assemble into anisotropic mesostructures. Strictly speaking, we know that the magnetostatic energy of an aggregate will be influenced by its shape, interfacial structures, and interior microstructure~\cite{conradt2026b,conradt2026}. However, because aggregates contain of the order $10^5$ particles and persist for as much as $10^5$ $\tau_D$, a direct Brownian dynamics simulation of mutually polarizing dipolar spheres is infeasible. We thus attempt to capture large scale trends by adapting the mean-field calculation of Grasselli \emph{et al.}~\cite{grasselli1994} and Promislow and Gast~\cite{promislow1997c}, representing a single aggregate as a triaxial ellipsoid. 

We start with the expression for the magnetization of a uniformly magnetized, isolated ellipsoid,
\begin{equation} \label{eq:3}
	\mathbf{M}_{a} = \mu_0 V_a \left[ \frac{\chi_a \mathbf{H}_0}{1+\chi_a n_z} \right].
\end{equation}
Here $\mu_0$ is the vacuum permeability, $V_a$ is the aggregate volume, $\mathbf{H}_0$ is the applied magnetic field, $\chi_a$ is the aggregate's average magnetic susceptibility, and $n_z$ is called the demagnetizing factor. The volume of an ellipsoid is defined by $V_a=(4/3)\pi a b c$ where $a$, $b$, and $c$ are the semi-axes which correspond to the half-lengths of the volume along its three axes of symmetry. The average magnetic susceptibility is calculated through a mean-field theory given by Bruggeman for conductive spherical inclusions in insulating media~\cite{promislow1997c},
\begin{equation}\label{eq:10}
	\frac{\chi_a-\chi_p}{\chi_m-\chi_p}\left(\frac{1+\chi_m}{1+\chi_a}\right)^{\frac{1}{3}}=1-\phi_a,
\end{equation}
where $\chi_p$ is the particle susceptibility and $\chi_m$ is the fluid susceptibility. The demagnetizing factor relates the geometry of the body to the magnitude of the depolarization field that resists magnetization of the body. Algebraic formulas for demagnetizing factors exist for ellipsoids of constrained geometries, for example the ellipsoid of revolution; for general triaxial ellipsoids an integral solution must be numerically evaluated~\cite{beleggia2006},
\begin{equation} \label{eq:4}
	n_{z}=\frac{1}{2}\int_{0}^{\infty} \frac{1}{\left(1+u\right)^{1.5}}\frac{1}{ \sqrt{1+\frac{u}{\gamma^{2}}}}\frac{1}{\sqrt{1+\frac{u}{\beta^{2}}}}\,\mathrm{d}u.
\end{equation}
The parameters $\gamma$ and $\beta$ are defined as the ratios of the minor axes to the major axis, $\gamma=c/a$ and $\beta=b/a$. Previous work has used algebraic approximations for the demagnetizing factor which restrict the values of $a$, $b$, and $c$ to those corresponding to ellipsoids of revolution or prolate ellipsoids. Using this expression, the demagnetizing factor can be evaluated for an ellipsoid of any shape. The demagnetizing factor varies from 0 to 1. As sphericity increases, $n_z$ increases, which increases the demagnetizing field (reducing the net magnetization), raises the aggregate energy, and thus favors elongation (smaller $n_z$). This elongation is mitigated by a surface energy which accounts for the decrease of the local fields at a surface resulting from lost interactions with absent neighbors at the surface. Our use of this integral form is a generalization of past work which assumes oblate or prolate ellipsoids and could not approximate the aggregate aspect ratios implied by our sample-spanning ribbons.

\begin{figure*}
\includegraphics[width=\textwidth]{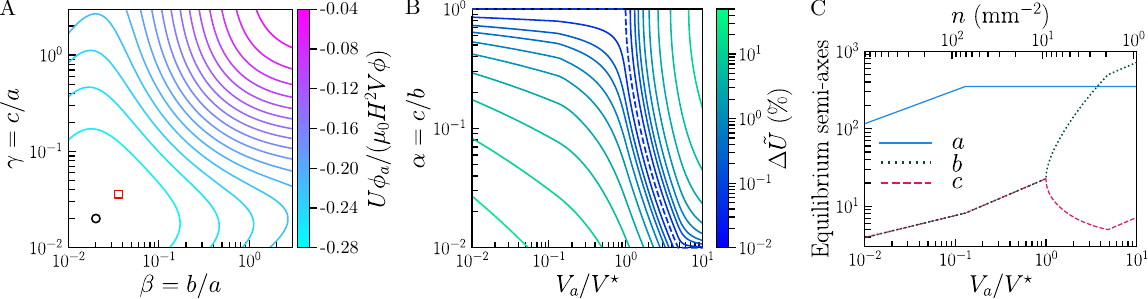}
\caption{(A) Single-aggregate energy nondimensionalized by the background magnetic energy density of the field is plotted as a function of ellipsoid shape parameters $\gamma$ and $\beta$. Surface thickness is taken as one particle diameter in thickness. The low-energy point is highlighted by a black circle. (B) Energy difference from the low-energy geometry as a function of normalized aggregate volume. Volumes are normalized by $V^{\star}$, the volume where the equilibrium $\alpha$ value taken by minimizing Eq.~\eqref{eq:result} deviates from one. $a$ corresponds to depth along the field axis and is bounded by $1\le a < 350$, while $b,c\ge 1$. The dotted blue line gives the low-energy aggregate elongation versus aggregate volume. (C) Ellipsoid dimensions $a$, $b$, and $c$ predicted for a confined aggregate plotted against $V_a/V^\star$ and the nominal aggregate number density. }
\label{fig:analysis}
\end{figure*}

The surface energy is accounted for through a surface correction term, $n_\sigma$, yielding a modified expression for the aggregate magnetization,
\begin{equation} \label{eq:3sigma}
	\mathbf{M}_{a} = \mu_0 V_a \left[ \frac{\chi_a \mathbf{H}_0}{1+ \chi_a n_z} \right]\left(1-n_\sigma \right). 
\end{equation}
In reality, the value of $n_\sigma$ is sensitive to surface orientation and shell thickness. Calculations of dipolar crystal surfaces, including recent mutually polarizing treatments, show strong dependence of surface energy on surface angle, crystal structure, particle polarizability, and local surface arrangement~\cite{toor1992,clercx1993,conradt2026}.  We use the approximation of  Promislow and Gast, considering that a fraction of the ellipsoid's volume exhibits a ``surface'' polarization consisting of one-half of the bulk polarization, corresponding to the reduction in the single-particle Lorentz field from $(1/3) M_a$ to $(1/6) M_a$. This yields~\cite{promislow1997c}
\begin{equation} \label{eq:8}
	n_{\sigma,p}=\frac{V_s}{V_a}\frac{\chi_p \phi_a}{6\chi_a},
\end{equation}
where $V_s$ and $V_a$ refer to the surface and aggregate volumes respectively, with the surface volume defined as $V_s=V_a-(4/3)\pi(a-\delta)(b-\delta)(c-\delta)$. This corresponds to an ellipsoidal shell concentric with the aggregate with thickness $\delta$ that is taken to be one particle diameter ($2R$) in depth. This value is chosen as several calculations of surface energies in polarized particle crystals have shown that surface energy contributions stem primarily from the first layer of particles~\cite{clercx1993}. The aggregate potential is additionally influenced by interactions with neighbors. The field perturbation caused by inter-aggregate interactions manifests as an additional factor in the expression for the aggregate magnetization,
\begin{equation} \label{eq:11}
	\mathbf{M}_a = \mathbf{M}_a^h \frac{1-n_\sigma}{1+n_r(1-n_\sigma)},
\end{equation}
where $\mathbf{M}_a^h$ is the aggregate magnetization in the absence of inter-aggregate interactions and $n_r$ is an interaction factor. Test calculations at physically observed number densities give $n_r \approx 0.10$, so we neglect this modest correction and use the resulting single-aggregate energy,
\begin{equation}
	\tilde{U}_a =  \frac{U_a \phi_a}{\mu_0 V_a H_0^2} = - \frac{1}{2} \left[ \frac{\chi_a}{1+ \chi_a n_z} \right] (1-n_\sigma).
\label{eq:result}
\end{equation}
Equation~\eqref{eq:result} gives a simple means to map the geometry of a self-assembled domain to its magnetostatic self-energy.

\subsection{Confinement and aggregate shape anisotropy}

We now apply our model towards explaining the suspension structures we see in the InSPACE-4 experiments. We take as physical parameters a fluid susceptibility of $\chi_m=0$ and a particle susceptibility of $\chi_p=1.4$, and we assume that the aggregate interior is a BCT lattice with $\phi_a=2\pi/9$. Using the effective-medium estimate for a BCT interior gives an effective aggregate susceptibility $\chi_a\simeq 0.88$. A surface shell of thickness $\delta=2R\approx 1~\upmu\mathrm{m}$ is used to evaluate Eq.~\eqref{eq:8}. These inputs specify the model for the particles used in microgravity, and are sufficient to apply Eq.~\eqref{eq:result} to aggregates in our microgravity experiments.

We consider the energy of an isolated aggregate and take $\tilde{U}(\beta,\gamma)$ to be the energy of an isolated aggregate as a function of $\beta=b/a$ and $\gamma=c/a$. Solving $\tilde{U}(\beta,\gamma)$ over a fine grid of $\{a,b,c\}$ yields an energy landscape illustrated in Fig.~\ref{fig:analysis}A. The only constraints on $\{a,b,c\}$ are set by the shell thickness $\delta$. We relate ellipsoid axes to the measured dimensions in Fig.~\ref{fig:lenswidths} by taking $a\sim D/(4R)$, $b\sim \langle L\rangle/(4R)$, and $c\sim \langle T\rangle/(4R)$, where $D$ is the aggregate depth along the field axis.  The surface is symmetric about $\beta=\gamma$, and the minimum occurs at an ellipsoid of revolution with $\beta=\gamma\approx 0.02$. Per the ST-view images examined in this study, this structure would correspond to a circular cross-section that extends $\sim 50$ thicknesses along the field axis, which translates to semi-axes of $a=513$ and $b=c=10.3$, marked by a black circle on Fig.~\ref{fig:analysis}A. However, in our system, the capillary walls force a limit $a\le 350$. Therefore, the lowest-energy accessible shape is an ellipsoid of revolution with $a=350$ and $b=c=12.5$, marked by a red square on Fig.~\ref{fig:analysis}A.
 
 \begin{figure*}
\includegraphics[width=5in]{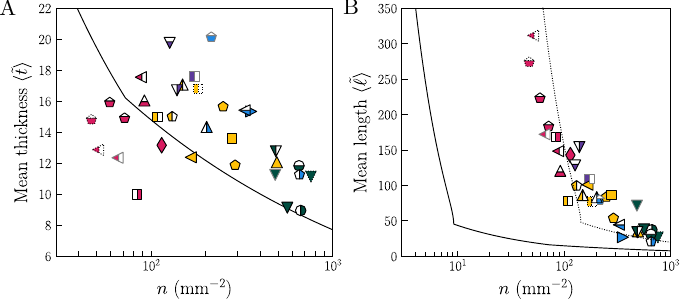}
\caption{Model predictions for the average dimensions of self-assembled colloidal aggregates versus experimental measurements. Predictions are obtained by minimizing the single-aggregate, time-averaged magnetostatic energy in Eq.~\eqref{eq:result} with respect to the ellipsoid axes $(a,b,c)$ where $a$ is the field-parallel (depth) axis. (A) Shows the experimentally measured $\langle \tilde{T}\rangle$ (markers) against the value $2c$ computed by minimizing Eq.~\eqref{eq:result} over $(a,b,c)$ subject to $a\leq350$. (B) gives the analogous comparison for the experimentally measured $\langle \tilde{L}\rangle$, which corresponds to $2b$. The dotted line represents the predicted thicknesses if the calculation assumed $\phi=0.06$. }
\label{fig:analysis2}
\end{figure*}

The low-energy aggregate shape is therefore isotropic in the plane perpendicular to the field, dissimilar to the structures we observe in the self-assembly experiments. We next consider whether it is plausible for aggregate volumes to exceed that of the lowest-energy, accessible geometry. The $a=350$, $b=c=12.5$ ellipsoid has volume $V^\star\approx 2.3\times 10^5~\upmu\mathrm{m}^3$. Assuming an interior particle volume fraction $\phi_a=2\pi/9$ and particle volume $V_p=(4/3)\pi R^3$, this corresponds to $N^\star\approx 3\times 10^5$ particles. The nominal particle count in the field of view is $N_t=\phi V_t/(\phi_a V_p)$, where $V_t$ is the estimated imaged volume. Taking $V_t\approx 3.8\times10^8~\upmu\mathrm{m}^3$ (a $700~\upmu\mathrm{m} \times 700~\upmu\mathrm{m}$ area and $\sim 700~\upmu\mathrm{m}$ depth) gives $N_t\approx 5\times10^6$. This implies that when the aggregate number density falls below of order $30$--$40~\mathrm{mm}^{-2}$, the equilibrium geometry predicted by Eq.~\eqref{eq:result} is inaccessible due to confinement. Aggregates with greater particle numbers must then either flatten anisotropically or thicken isotropically. The outcome is not obvious, because both the demagnetizing integral in Eq.~\eqref{eq:4} and the ratio of the surface shell volume to bulk volume in Eq.~\eqref{eq:8} depend nonlinearly on ellipsoid shape.

To determine the low-energy structure of confined aggregates, we reevaluate Eq.~\eqref{eq:result} under the constraint $1<a<350$ over aggregate volumes $V_a/V^\star\in[10^{-2},10]$. We parameterize flattening by $\alpha=\gamma/\beta=c/b$, where $\alpha<1$ indicates sheet- or ribbon-like cross-sections and $\alpha=1$ corresponds to a circular cross-section. We then compute $\tilde{U}_{\min}$ on a dense $2000\times 2000$ $\alpha$ versus $V_a/V^\star$ grid. The resulting surface (Fig.~\ref{fig:analysis}B) gives the $\alpha$ that minimizes the energy at each $V_a$ (dotted blue line), and the energetic penalty $\Delta\tilde{U}$ for departing from it. For $V_a\le V^\star$, the minimum occurs at $\alpha=1$ (isotropic cross-sections). For $V_a>V^\star$, the minimum shifts to $\alpha<1$, so growth beyond $V^\star$ favors anisotropic flattening. This trend is consistent with the images. As the number density of aggregates decreases, domains elongate and become more anisotropic (cf.\ Fig.~\ref{fig:lenswidths}). In this model, the preference for flattening perpendicular to the field arises from the competition between demagnetizing and surface energies under confinement.

We next directly compare the model prediction to measured aggregate dimensions. Fig.~\ref{fig:analysis}C summarizes the model’s equilibrium geometry as aggregate volume increases (or, equivalently, as the nominal number density decreases). Assuming the global particle volume fraction is conserved within the observed region, the ellipsoid depth along the field ($a$) and the in-plane axes ($b,c$) grow with power-law slopes of 0.43 and 0.20 respectively, maintaining an isotropic cross-section ($b=c$). The ellipsoid becomes confined when $a$ reaches its cap at approximately $0.1V^\star$; beyond this point $b$ and $c$ continue to increase together (isotropic in-plane thickening) with a power law slope of 0.5 until $V^\star$, where the isotropic cross-section becomes unfavorable and the axes separate. For $V>V^\star$, the model predicts a flattening transition where the length $b$ increases rapidly, potentially extending beyond the field of view, while the minor dimension $c$ decreases, corresponding to thinning. Beyond approximately $40V^\star$, the slopes change again and $b$ and $c$ resume power-law growth with a slope of 0.5 such that elongated aggregates both thicken and lengthen at comparable rates as volume increases.

The semiquantitative predictions of the model map well to the trends observed in our experiments: as aggregate number density decreases (i.e., volume increases), aggregates become more anisotropic. The model additionally makes quantitative predictions once a real length scale is fixed by the surface thickness. The predicted aggregate thicknesses (mapping to $2c$) and lengths (mapping to $2b$) are plotted against the measured averages $\langle \tilde{T}\rangle$ and $\langle \tilde{L}\rangle$ in Fig.~\ref{fig:analysis2}A and B, respectively. The agreement is mixed. Thicknesses follow the ellipsoidal prediction well; the model recovers both the magnitudes and the modest negative slope with increasing number density. For lengths, the model captures the sharp divergence characteristic of the ribbons and sheets phases, but it places the onset of that divergence at substantially lower nominal number densities (i.e., larger volumes) than we observed.

Several simplifications potentially limit the \emph{quantitative} accuracy of this description. The model uses a mean-field effective susceptibility, a triaxial-ellipsoid representation of aggregates, and an isotropic proxy for interfacial energy even though recent calculations show that local fields and surface energies depend on crystal structure, surface orientation, and are influenced by the mutual polarization of densely packed particle assemblies~\cite{conradt2026b,conradt2026,sherman2019c}. It also assumes the nominal particle volume fraction within the imaged region; deviations from that assumption would shift the curves in Fig.~\ref{fig:analysis2} without altering the qualitative prediction that confinement favors anisotropic flattening once aggregates grow beyond the isotropic confined state. 

Our analysis implies that confinement plays a role in the aggregate mesostructure. The mean-field ellipsoid model predicts anisotropic flattening with increasing aggregate volume, consistent with what we observe. It reproduces the minor dimension of self-assembled domains but places the onset of extreme elongation at lower nominal number density than we measured. From the kinetics, continued growth depends on phase; in the arrested and sheet states, aggregation largely stops after the initial onset of structure. One possibility is that cluster size is set during condensation and that further coarsening is limited when merging and splitting dynamics are weak. Even so, the model's successes indicate that the observed morphologies remain coupled to an underlying equilibrium magnetostatic energy landscape, while the dynamic spiky and ribbon phases also require genuinely dissipative mechanisms. How those two effects couple remains an open problem~\cite{sherman2016,kach2024}. 

\section{Conclusion}

Colloids of paramagnetic spheres under microgravity conditions exhibit phase separation when subjected to sufficiently strong, unidirectional magnetic fields. By applying a square-wave toggled field and varying parameters such as field strength, toggle frequency, and duty ratio, one can manipulate the suspension structure across multiple length scales. Under relatively low fields ($627\,\mathrm{A/m}$), the system remains disordered, showing no discernible aggregation. In the higher field regime ($1164$--$2276\,\mathrm{A/m}$), we observe five distinct structured morphologies: a dynamically arrested (gel-like) phase at small $t_{\mathrm{off}}$, a spiky phase and a sheets phase at intermediate $t_{\mathrm{off}}$, a highly elongated ribbons phase, and a slowly coarsening fluid-fluid regime at the largest $t_{\mathrm{off}}$. The resulting phase diagram is consistent with both analogous terrestrial experiments~\cite{kim2020} and Brownian dynamics simulations of dipolar colloids~\cite{sherman2019c}. Moreover, our mean-field analysis of the magnetic potential energy of aggregates indicates that confinement may promote the breaking of symmetry associated with emergent, anisotropic structures.

These findings have broader implications for designing and understanding field-responsive materials. In particular, the demagnetizing fields of large colloidal aggregates coupled with local field effects induced by particle-scale organization both affect the internal magnetic fields experienced by embedded colloidal matter. This suggests a route to tailoring magnetic properties by tuning the shapes and internal structures of colloidal assemblies. There is precedent for using self-assembly to manipulate the bulk magnetic properties of magnetic materials; for example, the phenomenon of ``giant susceptibility'' has been noted where paramagnetic nanoparticles field-directed to aggregate exhibit linear susceptibilities far in excess of their single-particle susceptibilities due to local and demagnetizing field effects~\cite{vanrhee2013,martin2008}. Future research could focus on quantitatively linking the aggregate shape with magnetization characteristics, opening the possibility of engineering magnetically tunable colloidal materials.

More broadly, this study offers a foundation for the analysis of dissipative self-assembly, wherein nonequilibrium steady states are created and maintained through dynamic processes rather than pure thermodynamic control. By mitigating the influence of gravitational fields, we present a more idealized example for such a process wherein we better isolate a relatively simple system---dilute, monodisperse paramagnetic colloidal particles subjected to an easily toggled and relatively simple external stimulus, a uniform external field. Although we have not provided a full theoretical mechanism for the emerging structures and their dynamics,  the quantitative insights presented here can serve as a foundational reference for interpreting and guiding experimental, computational, and theoretical work in the future.

\begin{acknowledgments}
The authors acknowledge funding from the National Science Foundation (CBET-1637991) and NASA (80NSSC24K0189){.} J.\ C.\ was supported in part by a GAANN Fellowship funded by the Department of Education (P200A210065). We also acknowledge helpful discussions with Z.\ Sherman.
\end{acknowledgments}

\section{ Supplemental Material }

The Supplemental Material includes:
\begin{itemize}
  \item A summary of the overall InSPACE-4 flight matrix, observation volume, and relation between the full campaign and the analyzed subset;
  \item An analysis of the geomagnetic field on the ISS and its relation to aggregate orientation and alignment;
  \item Calculations and simulations of the Helmholtz-coil field distribution, including field uniformity within the imaged region and axial variation along the capillary;
  \item Estimates of the ISS microgravity acceleration environment and its expected effect on sedimentation and inter-aggregate interactions;
  \item Additional kinetic analyses of structure evolution, including the time evolution of area coverage, aggregate number density, aggregate length and thickness, and a steady-state criterion based on transmitted-intensity drift; and
  \item Details of the morphological \(k\)-means classification used to define the steady-state phases, including the feature definitions and cluster-validity metrics.
\end{itemize}


\begin{thebibliography}{71}%
\makeatletter
\providecommand \@ifxundefined [1]{%
 \@ifx{#1\undefined}
}%
\providecommand \@ifnum [1]{%
 \ifnum #1\expandafter \@firstoftwo
 \else \expandafter \@secondoftwo
 \fi
}%
\providecommand \@ifx [1]{%
 \ifx #1\expandafter \@firstoftwo
 \else \expandafter \@secondoftwo
 \fi
}%
\providecommand \natexlab [1]{#1}%
\providecommand \enquote  [1]{``#1''}%
\providecommand \bibnamefont  [1]{#1}%
\providecommand \bibfnamefont [1]{#1}%
\providecommand \citenamefont [1]{#1}%
\providecommand \href@noop [0]{\@secondoftwo}%
\providecommand \href [0]{\begingroup \@sanitize@url \@href}%
\providecommand \@href[1]{\@@startlink{#1}\@@href}%
\providecommand \@@href[1]{\endgroup#1\@@endlink}%
\providecommand \@sanitize@url [0]{\catcode `\\12\catcode `\$12\catcode
  `\&12\catcode `\#12\catcode `\^12\catcode `\_12\catcode `\%12\relax}%
\providecommand \@@startlink[1]{}%
\providecommand \@@endlink[0]{}%
\providecommand \url  [0]{\begingroup\@sanitize@url \@url }%
\providecommand \@url [1]{\endgroup\@href {#1}{\urlprefix }}%
\providecommand \urlprefix  [0]{URL }%
\providecommand \Eprint [0]{\href }%
\providecommand \doibase [0]{https://doi.org/}%
\providecommand \selectlanguage [0]{\@gobble}%
\providecommand \bibinfo  [0]{\@secondoftwo}%
\providecommand \bibfield  [0]{\@secondoftwo}%
\providecommand \translation [1]{[#1]}%
\providecommand \BibitemOpen [0]{}%
\providecommand \bibitemStop [0]{}%
\providecommand \bibitemNoStop [0]{.\EOS\space}%
\providecommand \EOS [0]{\spacefactor3000\relax}%
\providecommand \BibitemShut  [1]{\csname bibitem#1\endcsname}%
\let\auto@bib@innerbib\@empty
\bibitem [{\citenamefont {Pusey}\ and\ \citenamefont {{van
  Megen}}(1986)}]{Pusey1986}%
  \BibitemOpen
  \bibfield  {author} {\bibinfo {author} {\bibfnamefont {P.~N.}\ \bibnamefont
  {Pusey}}\ and\ \bibinfo {author} {\bibfnamefont {W.}~\bibnamefont {{van
  Megen}}},\ }\bibfield  {title} {\bibinfo {title} {Phase-behavior of
  concentrated suspensions of nearly hard colloidal spheres},\ }\href@noop {}
  {\bibfield  {journal} {\bibinfo  {journal} {Nature}\ }\textbf {\bibinfo
  {volume} {320}},\ \bibinfo {pages} {340} (\bibinfo {year}
  {1986})}\BibitemShut {NoStop}%
\bibitem [{\citenamefont {Zhu}\ \emph {et~al.}(1997)\citenamefont {Zhu},
  \citenamefont {Li}, \citenamefont {Rogers}, \citenamefont {Meyer},
  \citenamefont {Ottewill}, \citenamefont {Crew}, \citenamefont {Russel},\ and\
  \citenamefont {Chaikin}}]{Zhu1997b}%
  \BibitemOpen
  \bibfield  {author} {\bibinfo {author} {\bibfnamefont {J.}~\bibnamefont
  {Zhu}}, \bibinfo {author} {\bibfnamefont {M.}~\bibnamefont {Li}}, \bibinfo
  {author} {\bibfnamefont {R.}~\bibnamefont {Rogers}}, \bibinfo {author}
  {\bibfnamefont {W.}~\bibnamefont {Meyer}}, \bibinfo {author} {\bibfnamefont
  {R.~H.}\ \bibnamefont {Ottewill}}, \bibinfo {author} {\bibfnamefont {S.-.
  S.~S.}\ \bibnamefont {Crew}}, \bibinfo {author} {\bibfnamefont {W.~B.}\
  \bibnamefont {Russel}},\ and\ \bibinfo {author} {\bibfnamefont {P.~M.}\
  \bibnamefont {Chaikin}},\ }\bibfield  {title} {\bibinfo {title}
  {Crystallization of hard-sphere colloids in microgravity},\ }\href@noop {}
  {\bibfield  {journal} {\bibinfo  {journal} {Nature}\ }\textbf {\bibinfo
  {volume} {387}},\ \bibinfo {pages} {883} (\bibinfo {year}
  {1997})}\BibitemShut {NoStop}%
\bibitem [{\citenamefont {Yethiraj}\ and\ \citenamefont
  {Van~Blaaderen}(2003)}]{yethiraj2003}%
  \BibitemOpen
  \bibfield  {author} {\bibinfo {author} {\bibfnamefont {A.}~\bibnamefont
  {Yethiraj}}\ and\ \bibinfo {author} {\bibfnamefont {A.}~\bibnamefont
  {Van~Blaaderen}},\ }\bibfield  {title} {\bibinfo {title} {A colloidal model
  system with an interaction tunable from hard sphere to soft and dipolar},\
  }\href {https://doi.org/10.1038/nature01328} {\bibfield  {journal} {\bibinfo
  {journal} {Nature}\ }\textbf {\bibinfo {volume} {421}},\ \bibinfo {pages}
  {513} (\bibinfo {year} {2003})}\BibitemShut {NoStop}%
\bibitem [{\citenamefont {Lumsdon}\ \emph {et~al.}(2004)\citenamefont
  {Lumsdon}, \citenamefont {Kaler},\ and\ \citenamefont {Velev}}]{Lumsdon2004}%
  \BibitemOpen
  \bibfield  {author} {\bibinfo {author} {\bibfnamefont {S.~O.}\ \bibnamefont
  {Lumsdon}}, \bibinfo {author} {\bibfnamefont {E.~W.}\ \bibnamefont {Kaler}},\
  and\ \bibinfo {author} {\bibfnamefont {O.~D.}\ \bibnamefont {Velev}},\
  }\bibfield  {title} {\bibinfo {title} {Two-dimensional crystallization of
  microspheres by a coplanar {{AC}} electric field},\ }\href@noop {} {\bibfield
   {journal} {\bibinfo  {journal} {Langmuir}\ }\textbf {\bibinfo {volume}
  {20}},\ \bibinfo {pages} {2108} (\bibinfo {year} {2004})}\BibitemShut
  {NoStop}%
\bibitem [{\citenamefont {Ma}\ \emph {et~al.}(2013)\citenamefont {Ma},
  \citenamefont {Wu},\ and\ \citenamefont {Wu}}]{ma2013}%
  \BibitemOpen
  \bibfield  {author} {\bibinfo {author} {\bibfnamefont {F.}~\bibnamefont
  {Ma}}, \bibinfo {author} {\bibfnamefont {D.~T.}\ \bibnamefont {Wu}},\ and\
  \bibinfo {author} {\bibfnamefont {N.}~\bibnamefont {Wu}},\ }\bibfield
  {title} {\bibinfo {title} {Formation of colloidal molecules induced by
  alternating-current electric fields},\ }\href
  {https://doi.org/10.1021/ja403172p} {\bibfield  {journal} {\bibinfo
  {journal} {Journal of the American Chemical Society}\ }\textbf {\bibinfo
  {volume} {135}},\ \bibinfo {pages} {7839} (\bibinfo {year}
  {2013})}\BibitemShut {NoStop}%
\bibitem [{\citenamefont {Du}\ \emph {et~al.}(2013)\citenamefont {Du},
  \citenamefont {Li}, \citenamefont {Thakur},\ and\ \citenamefont
  {Biswal}}]{Du2013}%
  \BibitemOpen
  \bibfield  {author} {\bibinfo {author} {\bibfnamefont {D.}~\bibnamefont
  {Du}}, \bibinfo {author} {\bibfnamefont {D.}~\bibnamefont {Li}}, \bibinfo
  {author} {\bibfnamefont {M.}~\bibnamefont {Thakur}},\ and\ \bibinfo {author}
  {\bibfnamefont {S.~L.}\ \bibnamefont {Biswal}},\ }\bibfield  {title}
  {\bibinfo {title} {Generating an in situ tunable interaction potential for
  probing 2-{{D}} colloidal phase behavior},\ }\href
  {https://doi.org/10.1039/c3sm27620a} {\bibfield  {journal} {\bibinfo
  {journal} {Soft Matter}\ }\textbf {\bibinfo {volume} {9}},\ \bibinfo {pages}
  {6867} (\bibinfo {year} {2013})}\BibitemShut {NoStop}%
\bibitem [{\citenamefont {Gast}\ \emph {et~al.}(1983)\citenamefont {Gast},
  \citenamefont {Hall},\ and\ \citenamefont {Russel}}]{Gast1983}%
  \BibitemOpen
  \bibfield  {author} {\bibinfo {author} {\bibfnamefont {A.~P.}\ \bibnamefont
  {Gast}}, \bibinfo {author} {\bibfnamefont {C.~K.}\ \bibnamefont {Hall}},\
  and\ \bibinfo {author} {\bibfnamefont {W.~B.}\ \bibnamefont {Russel}},\
  }\bibfield  {title} {\bibinfo {title} {Polymer-induced phase separation in
  non-aqueous colloidal suspensions},\ }\href@noop {} {\bibfield  {journal}
  {\bibinfo  {journal} {J Colloid Interface Sci.}\ }\textbf {\bibinfo {volume}
  {96}},\ \bibinfo {pages} {251} (\bibinfo {year} {1983})}\BibitemShut
  {NoStop}%
\bibitem [{\citenamefont {He}\ \emph {et~al.}(2020)\citenamefont {He},
  \citenamefont {Gales}, \citenamefont {Ducrot}, \citenamefont {Gong},
  \citenamefont {Yi}, \citenamefont {Sacanna},\ and\ \citenamefont
  {Pine}}]{He2020}%
  \BibitemOpen
  \bibfield  {author} {\bibinfo {author} {\bibfnamefont {M.}~\bibnamefont
  {He}}, \bibinfo {author} {\bibfnamefont {J.~P.}\ \bibnamefont {Gales}},
  \bibinfo {author} {\bibfnamefont {{\'E}.}~\bibnamefont {Ducrot}}, \bibinfo
  {author} {\bibfnamefont {Z.}~\bibnamefont {Gong}}, \bibinfo {author}
  {\bibfnamefont {G.-R.}\ \bibnamefont {Yi}}, \bibinfo {author} {\bibfnamefont
  {S.}~\bibnamefont {Sacanna}},\ and\ \bibinfo {author} {\bibfnamefont {D.~J.}\
  \bibnamefont {Pine}},\ }\bibfield  {title} {\bibinfo {title} {Colloidal
  diamond},\ }\href {https://doi.org/10.1038/s41586-020-2718-6} {\bibfield
  {journal} {\bibinfo  {journal} {Nature}\ }\textbf {\bibinfo {volume} {585}},\
  \bibinfo {pages} {524} (\bibinfo {year} {2020})}\BibitemShut {NoStop}%
\bibitem [{\citenamefont {Ofosu}\ \emph {et~al.}(2024)\citenamefont {Ofosu},
  \citenamefont {Wilcoxson}, \citenamefont {Lee}, \citenamefont {Brackett},
  \citenamefont {Choi}, \citenamefont {Truskett},\ and\ \citenamefont
  {Milliron}}]{ofosu2024}%
  \BibitemOpen
  \bibfield  {author} {\bibinfo {author} {\bibfnamefont {C.~K.}\ \bibnamefont
  {Ofosu}}, \bibinfo {author} {\bibfnamefont {T.}~\bibnamefont {Wilcoxson}},
  \bibinfo {author} {\bibfnamefont {T.-L.}\ \bibnamefont {Lee}}, \bibinfo
  {author} {\bibfnamefont {W.}~\bibnamefont {Brackett}}, \bibinfo {author}
  {\bibfnamefont {J.}~\bibnamefont {Choi}}, \bibinfo {author} {\bibfnamefont
  {T.}~\bibnamefont {Truskett}},\ and\ \bibinfo {author} {\bibfnamefont
  {D.}~\bibnamefont {Milliron}},\ }\href
  {https://doi.org/10.26434/chemrxiv-2024-4r7fz} {\bibinfo {title} {Assessing
  {{Depletion Attractions Between Colloidal Nanocrystals}}}} (\bibinfo {year}
  {2024})\BibitemShut {NoStop}%
\bibitem [{\citenamefont {Fialkowski}\ \emph {et~al.}(2006)\citenamefont
  {Fialkowski}, \citenamefont {Bishop}, \citenamefont {Klajn}, \citenamefont
  {Smoukov}, \citenamefont {Campbell},\ and\ \citenamefont
  {Grzybowski}}]{fialkowski2006}%
  \BibitemOpen
  \bibfield  {author} {\bibinfo {author} {\bibfnamefont {M.}~\bibnamefont
  {Fialkowski}}, \bibinfo {author} {\bibfnamefont {K.~J.~M.}\ \bibnamefont
  {Bishop}}, \bibinfo {author} {\bibfnamefont {R.}~\bibnamefont {Klajn}},
  \bibinfo {author} {\bibfnamefont {S.~K.}\ \bibnamefont {Smoukov}}, \bibinfo
  {author} {\bibfnamefont {C.~J.}\ \bibnamefont {Campbell}},\ and\ \bibinfo
  {author} {\bibfnamefont {B.~A.}\ \bibnamefont {Grzybowski}},\ }\bibfield
  {title} {\bibinfo {title} {Principles and {{Implementations}} of
  {{Dissipative}} ({{Dynamic}}) {{Self-Assembly}}},\ }\href
  {https://doi.org/10.1021/jp054153q} {\bibfield  {journal} {\bibinfo
  {journal} {The Journal of Physical Chemistry B}\ }\textbf {\bibinfo {volume}
  {110}},\ \bibinfo {pages} {2482} (\bibinfo {year} {2006})}\BibitemShut
  {NoStop}%
\bibitem [{\citenamefont {Grzelczak}\ \emph {et~al.}(2010)\citenamefont
  {Grzelczak}, \citenamefont {Vermant}, \citenamefont {Furst},\ and\
  \citenamefont {{Liz-Marz{\'a}n}}}]{grzelczak2010b}%
  \BibitemOpen
  \bibfield  {author} {\bibinfo {author} {\bibfnamefont {M.}~\bibnamefont
  {Grzelczak}}, \bibinfo {author} {\bibfnamefont {J.}~\bibnamefont {Vermant}},
  \bibinfo {author} {\bibfnamefont {E.~M.}\ \bibnamefont {Furst}},\ and\
  \bibinfo {author} {\bibfnamefont {L.~M.}\ \bibnamefont {{Liz-Marz{\'a}n}}},\
  }\bibfield  {title} {\bibinfo {title} {Directed {{Self-Assembly}} of
  {{Nanoparticles}}},\ }\href {https://doi.org/10.1021/nn100869j} {\bibfield
  {journal} {\bibinfo  {journal} {ACS Nano}\ }\textbf {\bibinfo {volume} {4}},\
  \bibinfo {pages} {3591} (\bibinfo {year} {2010})}\BibitemShut {NoStop}%
\bibitem [{\citenamefont {Forster}\ \emph {et~al.}(2011)\citenamefont
  {Forster}, \citenamefont {Park}, \citenamefont {Mittal}, \citenamefont {Noh},
  \citenamefont {Schreck}, \citenamefont {O'Hern}, \citenamefont {Cao},
  \citenamefont {Furst},\ and\ \citenamefont {Dufresne}}]{forster2011}%
  \BibitemOpen
  \bibfield  {author} {\bibinfo {author} {\bibfnamefont {J.~D.}\ \bibnamefont
  {Forster}}, \bibinfo {author} {\bibfnamefont {J.-G.}\ \bibnamefont {Park}},
  \bibinfo {author} {\bibfnamefont {M.}~\bibnamefont {Mittal}}, \bibinfo
  {author} {\bibfnamefont {H.}~\bibnamefont {Noh}}, \bibinfo {author}
  {\bibfnamefont {C.~F.}\ \bibnamefont {Schreck}}, \bibinfo {author}
  {\bibfnamefont {C.~S.}\ \bibnamefont {O'Hern}}, \bibinfo {author}
  {\bibfnamefont {H.}~\bibnamefont {Cao}}, \bibinfo {author} {\bibfnamefont
  {E.~M.}\ \bibnamefont {Furst}},\ and\ \bibinfo {author} {\bibfnamefont
  {E.~R.}\ \bibnamefont {Dufresne}},\ }\bibfield  {title} {\bibinfo {title}
  {Assembly of {{Optical-Scale Dumbbells}} into {{Dense Photonic Crystals}}},\
  }\href {https://doi.org/10.1021/nn202227f} {\bibfield  {journal} {\bibinfo
  {journal} {ACS Nano}\ }\textbf {\bibinfo {volume} {5}},\ \bibinfo {pages}
  {6695} (\bibinfo {year} {2011})}\BibitemShut {NoStop}%
\bibitem [{\citenamefont {Tagliazucchi}\ and\ \citenamefont
  {Szleifer}(2016)}]{tagliazucchi2016}%
  \BibitemOpen
  \bibfield  {author} {\bibinfo {author} {\bibfnamefont {M.}~\bibnamefont
  {Tagliazucchi}}\ and\ \bibinfo {author} {\bibfnamefont {I.}~\bibnamefont
  {Szleifer}},\ }\bibfield  {title} {\bibinfo {title} {Dynamics of dissipative
  self-assembly of particles interacting through oscillatory forces},\ }\href
  {https://doi.org/10.1039/C5FD00115C} {\bibfield  {journal} {\bibinfo
  {journal} {Faraday Discussions}\ }\textbf {\bibinfo {volume} {186}},\
  \bibinfo {pages} {399} (\bibinfo {year} {2016})}\BibitemShut {NoStop}%
\bibitem [{\citenamefont {{Arango-Restrepo}}\ \emph {et~al.}(2019)\citenamefont
  {{Arango-Restrepo}}, \citenamefont {Barrag{\'a}n},\ and\ \citenamefont
  {Rubi}}]{arango-restrepo2019}%
  \BibitemOpen
  \bibfield  {author} {\bibinfo {author} {\bibfnamefont {A.}~\bibnamefont
  {{Arango-Restrepo}}}, \bibinfo {author} {\bibfnamefont {D.}~\bibnamefont
  {Barrag{\'a}n}},\ and\ \bibinfo {author} {\bibfnamefont {J.~M.}\ \bibnamefont
  {Rubi}},\ }\bibfield  {title} {\bibinfo {title} {Self-assembling outside
  equilibrium: Emergence of structures mediated by dissipation},\ }\href
  {https://doi.org/10.1039/C9CP01088B} {\bibfield  {journal} {\bibinfo
  {journal} {Physical Chemistry Chemical Physics}\ }\textbf {\bibinfo {volume}
  {21}},\ \bibinfo {pages} {17475} (\bibinfo {year} {2019})}\BibitemShut
  {NoStop}%
\bibitem [{\citenamefont {Liljestr{\"o}m}\ \emph {et~al.}(2019)\citenamefont
  {Liljestr{\"o}m}, \citenamefont {Chen}, \citenamefont {Dommersnes},
  \citenamefont {Fossum},\ and\ \citenamefont {Gr{\"o}schel}}]{liljestrom2019}%
  \BibitemOpen
  \bibfield  {author} {\bibinfo {author} {\bibfnamefont {V.}~\bibnamefont
  {Liljestr{\"o}m}}, \bibinfo {author} {\bibfnamefont {C.}~\bibnamefont
  {Chen}}, \bibinfo {author} {\bibfnamefont {P.}~\bibnamefont {Dommersnes}},
  \bibinfo {author} {\bibfnamefont {J.~O.}\ \bibnamefont {Fossum}},\ and\
  \bibinfo {author} {\bibfnamefont {A.~H.}\ \bibnamefont {Gr{\"o}schel}},\
  }\bibfield  {title} {\bibinfo {title} {Active structuring of colloids through
  field-driven self-assembly},\ }\href
  {https://doi.org/10.1016/j.cocis.2018.10.008} {\bibfield  {journal} {\bibinfo
   {journal} {Current Opinion in Colloid \& Interface Science}\ }\textbf
  {\bibinfo {volume} {40}},\ \bibinfo {pages} {25} (\bibinfo {year}
  {2019})}\BibitemShut {NoStop}%
\bibitem [{\citenamefont {Coughlan}\ \emph {et~al.}(2019)\citenamefont
  {Coughlan}, \citenamefont {{Torres-D{\'\i}az}}, \citenamefont {Zhang},\ and\
  \citenamefont {Bevan}}]{coughlan2019}%
  \BibitemOpen
  \bibfield  {author} {\bibinfo {author} {\bibfnamefont {A.~C.~H.}\
  \bibnamefont {Coughlan}}, \bibinfo {author} {\bibfnamefont {I.}~\bibnamefont
  {{Torres-D{\'\i}az}}}, \bibinfo {author} {\bibfnamefont {J.}~\bibnamefont
  {Zhang}},\ and\ \bibinfo {author} {\bibfnamefont {M.~A.}\ \bibnamefont
  {Bevan}},\ }\bibfield  {title} {\bibinfo {title} {Non-equilibrium
  steady-state colloidal assembly dynamics},\ }\href
  {https://doi.org/10.1063/1.5094554} {\bibfield  {journal} {\bibinfo
  {journal} {The Journal of Chemical Physics}\ }\textbf {\bibinfo {volume}
  {150}},\ \bibinfo {pages} {204902} (\bibinfo {year} {2019})}\BibitemShut
  {NoStop}%
\bibitem [{\citenamefont {Sanchez}\ \emph {et~al.}(2012)\citenamefont
  {Sanchez}, \citenamefont {Chen}, \citenamefont {DeCamp}, \citenamefont
  {Heymann},\ and\ \citenamefont {Dogic}}]{sanchez2012}%
  \BibitemOpen
  \bibfield  {author} {\bibinfo {author} {\bibfnamefont {T.}~\bibnamefont
  {Sanchez}}, \bibinfo {author} {\bibfnamefont {D.~T.~N.}\ \bibnamefont
  {Chen}}, \bibinfo {author} {\bibfnamefont {S.~J.}\ \bibnamefont {DeCamp}},
  \bibinfo {author} {\bibfnamefont {M.}~\bibnamefont {Heymann}},\ and\ \bibinfo
  {author} {\bibfnamefont {Z.}~\bibnamefont {Dogic}},\ }\bibfield  {title}
  {\bibinfo {title} {Spontaneous motion in hierarchically assembled active
  matter},\ }\href {https://doi.org/10.1038/nature11591} {\bibfield  {journal}
  {\bibinfo  {journal} {Nature}\ }\textbf {\bibinfo {volume} {491}},\ \bibinfo
  {pages} {431} (\bibinfo {year} {2012})}\BibitemShut {NoStop}%
\bibitem [{\citenamefont {Cates}\ and\ \citenamefont
  {Tailleur}(2015)}]{cates2015}%
  \BibitemOpen
  \bibfield  {author} {\bibinfo {author} {\bibfnamefont {M.~E.}\ \bibnamefont
  {Cates}}\ and\ \bibinfo {author} {\bibfnamefont {J.}~\bibnamefont
  {Tailleur}},\ }\bibfield  {title} {\bibinfo {title} {Motility-{{Induced Phase
  Separation}}},\ }\href
  {https://doi.org/10.1146/annurev-conmatphys-031214-014710} {\bibfield
  {journal} {\bibinfo  {journal} {Annual Review of Condensed Matter Physics}\
  }\textbf {\bibinfo {volume} {6}},\ \bibinfo {pages} {219} (\bibinfo {year}
  {2015})}\BibitemShut {NoStop}%
\bibitem [{\citenamefont {Bishop}\ \emph {et~al.}(2023)\citenamefont {Bishop},
  \citenamefont {Biswal},\ and\ \citenamefont {Bharti}}]{bishop2023}%
  \BibitemOpen
  \bibfield  {author} {\bibinfo {author} {\bibfnamefont {K.~J.}\ \bibnamefont
  {Bishop}}, \bibinfo {author} {\bibfnamefont {S.~L.}\ \bibnamefont {Biswal}},\
  and\ \bibinfo {author} {\bibfnamefont {B.}~\bibnamefont {Bharti}},\
  }\bibfield  {title} {\bibinfo {title} {Active {{Colloids}} as {{Models}},
  {{Materials}}, and {{Machines}}},\ }\href
  {https://doi.org/10.1146/annurev-chembioeng-101121-084939} {\bibfield
  {journal} {\bibinfo  {journal} {Annual Review of Chemical and Biomolecular
  Engineering}\ }\textbf {\bibinfo {volume} {14}},\ \bibinfo {pages} {1}
  (\bibinfo {year} {2023})}\BibitemShut {NoStop}%
\bibitem [{\citenamefont {Martin}\ and\ \citenamefont
  {Snezhko}(2013)}]{martin2013}%
  \BibitemOpen
  \bibfield  {author} {\bibinfo {author} {\bibfnamefont {J.~E.}\ \bibnamefont
  {Martin}}\ and\ \bibinfo {author} {\bibfnamefont {A.}~\bibnamefont
  {Snezhko}},\ }\bibfield  {title} {\bibinfo {title} {Driving self-assembly and
  emergent dynamics in colloidal suspensions by time-dependent magnetic
  fields},\ }\href {https://doi.org/10.1088/0034-4885/76/12/126601} {\bibfield
  {journal} {\bibinfo  {journal} {Reports on Progress in Physics}\ }\textbf
  {\bibinfo {volume} {76}},\ \bibinfo {pages} {126601} (\bibinfo {year}
  {2013})}\BibitemShut {NoStop}%
\bibitem [{\citenamefont {{Spatafora-Salazar}}\ \emph
  {et~al.}(2021)\citenamefont {{Spatafora-Salazar}}, \citenamefont {Lobmeyer},
  \citenamefont {Cunha}, \citenamefont {Joshi},\ and\ \citenamefont
  {Biswal}}]{spatafora-salazar2021}%
  \BibitemOpen
  \bibfield  {author} {\bibinfo {author} {\bibfnamefont {A.}~\bibnamefont
  {{Spatafora-Salazar}}}, \bibinfo {author} {\bibfnamefont {D.~M.}\
  \bibnamefont {Lobmeyer}}, \bibinfo {author} {\bibfnamefont {L.~H.~P.}\
  \bibnamefont {Cunha}}, \bibinfo {author} {\bibfnamefont {K.}~\bibnamefont
  {Joshi}},\ and\ \bibinfo {author} {\bibfnamefont {S.~L.}\ \bibnamefont
  {Biswal}},\ }\bibfield  {title} {\bibinfo {title} {Hierarchical assemblies of
  superparamagnetic colloids in time-varying magnetic fields},\ }\href
  {https://doi.org/10.1039/D0SM01878C} {\bibfield  {journal} {\bibinfo
  {journal} {Soft Matter}\ }\textbf {\bibinfo {volume} {17}},\ \bibinfo {pages}
  {1120} (\bibinfo {year} {2021})}\BibitemShut {NoStop}%
\bibitem [{\citenamefont {Swan}\ \emph {et~al.}(2014)\citenamefont {Swan},
  \citenamefont {Vasquez},\ and\ \citenamefont {Furst}}]{swan2014}%
  \BibitemOpen
  \bibfield  {author} {\bibinfo {author} {\bibfnamefont {J.~W.}\ \bibnamefont
  {Swan}}, \bibinfo {author} {\bibfnamefont {P.~A.}\ \bibnamefont {Vasquez}},\
  and\ \bibinfo {author} {\bibfnamefont {E.~M.}\ \bibnamefont {Furst}},\
  }\bibfield  {title} {\bibinfo {title} {Buckling {{Instability}} of
  {{Self-Assembled Colloidal Columns}}},\ }\href
  {https://doi.org/10.1103/PhysRevLett.113.138301} {\bibfield  {journal}
  {\bibinfo  {journal} {Physical Review Letters}\ }\textbf {\bibinfo {volume}
  {113}},\ \bibinfo {pages} {138301} (\bibinfo {year} {2014})}\BibitemShut
  {NoStop}%
\bibitem [{\citenamefont {Tao}\ and\ \citenamefont {Sun}(1991)}]{tao1991}%
  \BibitemOpen
  \bibfield  {author} {\bibinfo {author} {\bibfnamefont {R.}~\bibnamefont
  {Tao}}\ and\ \bibinfo {author} {\bibfnamefont {J.~M.}\ \bibnamefont {Sun}},\
  }\bibfield  {title} {\bibinfo {title} {Three-dimensional structure of induced
  electrorheological solid},\ }\href
  {https://doi.org/10.1103/PhysRevLett.67.398} {\bibfield  {journal} {\bibinfo
  {journal} {Physical Review Letters}\ }\textbf {\bibinfo {volume} {67}},\
  \bibinfo {pages} {398} (\bibinfo {year} {1991})}\BibitemShut {NoStop}%
\bibitem [{\citenamefont {Toor}\ and\ \citenamefont {Halsey}(1992)}]{toor1992}%
  \BibitemOpen
  \bibfield  {author} {\bibinfo {author} {\bibfnamefont {W.~R.}\ \bibnamefont
  {Toor}}\ and\ \bibinfo {author} {\bibfnamefont {T.~C.}\ \bibnamefont
  {Halsey}},\ }\bibfield  {title} {\bibinfo {title} {Surface and bulk energies
  of dipolar lattices},\ }\href {https://doi.org/10.1103/PhysRevA.45.8617}
  {\bibfield  {journal} {\bibinfo  {journal} {Physical Review A}\ }\textbf
  {\bibinfo {volume} {45}},\ \bibinfo {pages} {8617} (\bibinfo {year}
  {1992})}\BibitemShut {NoStop}%
\bibitem [{\citenamefont {Sherman}(2019)}]{Sherman2019}%
  \BibitemOpen
  \bibfield  {author} {\bibinfo {author} {\bibfnamefont {Z.}~\bibnamefont
  {Sherman}},\ }\emph {\bibinfo {title} {Self-Assembly and Dynamics of
  Colloidal Dispersions in Steady and Time-Varying External Fields}},\
  \href@noop {} {Ph.D. thesis},\ \bibinfo  {school} {Massachusetts Institute of
  Technology. Department of Chemical Engineering} (\bibinfo {year}
  {2019})\BibitemShut {NoStop}%
\bibitem [{\citenamefont {Pal}\ \emph {et~al.}(2015)\citenamefont {Pal},
  \citenamefont {Malik}, \citenamefont {He}, \citenamefont {Ern{\'e}},
  \citenamefont {Yin}, \citenamefont {Kegel},\ and\ \citenamefont
  {Petukhov}}]{pal2015}%
  \BibitemOpen
  \bibfield  {author} {\bibinfo {author} {\bibfnamefont {A.}~\bibnamefont
  {Pal}}, \bibinfo {author} {\bibfnamefont {V.}~\bibnamefont {Malik}}, \bibinfo
  {author} {\bibfnamefont {L.}~\bibnamefont {He}}, \bibinfo {author}
  {\bibfnamefont {B.~H.}\ \bibnamefont {Ern{\'e}}}, \bibinfo {author}
  {\bibfnamefont {Y.}~\bibnamefont {Yin}}, \bibinfo {author} {\bibfnamefont
  {W.~K.}\ \bibnamefont {Kegel}},\ and\ \bibinfo {author} {\bibfnamefont
  {A.~V.}\ \bibnamefont {Petukhov}},\ }\bibfield  {title} {\bibinfo {title}
  {Tuning the {{Colloidal Crystal Structure}} of {{Magnetic Particles}} by
  {{External Field}}},\ }\href {https://doi.org/10.1002/anie.201409878}
  {\bibfield  {journal} {\bibinfo  {journal} {Angewandte Chemie International
  Edition}\ }\textbf {\bibinfo {volume} {54}},\ \bibinfo {pages} {1803}
  (\bibinfo {year} {2015})}\BibitemShut {NoStop}%
\bibitem [{\citenamefont {Promislow}\ and\ \citenamefont
  {Gast}(1996)}]{promislow1996}%
  \BibitemOpen
  \bibfield  {author} {\bibinfo {author} {\bibfnamefont {J.~H.~E.}\
  \bibnamefont {Promislow}}\ and\ \bibinfo {author} {\bibfnamefont {A.~P.}\
  \bibnamefont {Gast}},\ }\bibfield  {title} {\bibinfo {title}
  {Magnetorheological {{Fluid Structure}} in a {{Pulsed Magnetic Field}}},\
  }\href {https://doi.org/10.1021/la960104g} {\bibfield  {journal} {\bibinfo
  {journal} {Langmuir}\ }\textbf {\bibinfo {volume} {12}},\ \bibinfo {pages}
  {4095} (\bibinfo {year} {1996})}\BibitemShut {NoStop}%
\bibitem [{\citenamefont {Swan}\ \emph {et~al.}(2012)\citenamefont {Swan},
  \citenamefont {Vasquez}, \citenamefont {Whitson}, \citenamefont {Fincke},
  \citenamefont {Wakata}, \citenamefont {Magnus}, \citenamefont {Winne},
  \citenamefont {Barratt}, \citenamefont {Agui}, \citenamefont {Green},
  \citenamefont {Hall}, \citenamefont {Bohman}, \citenamefont {Bunnell},
  \citenamefont {Gast},\ and\ \citenamefont {Furst}}]{swan2012}%
  \BibitemOpen
  \bibfield  {author} {\bibinfo {author} {\bibfnamefont {J.~W.}\ \bibnamefont
  {Swan}}, \bibinfo {author} {\bibfnamefont {P.~A.}\ \bibnamefont {Vasquez}},
  \bibinfo {author} {\bibfnamefont {P.~A.}\ \bibnamefont {Whitson}}, \bibinfo
  {author} {\bibfnamefont {E.~M.}\ \bibnamefont {Fincke}}, \bibinfo {author}
  {\bibfnamefont {K.}~\bibnamefont {Wakata}}, \bibinfo {author} {\bibfnamefont
  {S.~H.}\ \bibnamefont {Magnus}}, \bibinfo {author} {\bibfnamefont {F.~D.}\
  \bibnamefont {Winne}}, \bibinfo {author} {\bibfnamefont {M.~R.}\ \bibnamefont
  {Barratt}}, \bibinfo {author} {\bibfnamefont {J.~H.}\ \bibnamefont {Agui}},
  \bibinfo {author} {\bibfnamefont {R.~D.}\ \bibnamefont {Green}}, \bibinfo
  {author} {\bibfnamefont {N.~R.}\ \bibnamefont {Hall}}, \bibinfo {author}
  {\bibfnamefont {D.~Y.}\ \bibnamefont {Bohman}}, \bibinfo {author}
  {\bibfnamefont {C.~T.}\ \bibnamefont {Bunnell}}, \bibinfo {author}
  {\bibfnamefont {A.~P.}\ \bibnamefont {Gast}},\ and\ \bibinfo {author}
  {\bibfnamefont {E.~M.}\ \bibnamefont {Furst}},\ }\bibfield  {title} {\bibinfo
  {title} {Multi-scale kinetics of a field-directed colloidal phase
  transition},\ }\href {https://doi.org/10.1073/pnas.1206915109} {\bibfield
  {journal} {\bibinfo  {journal} {Proceedings of the National Academy of
  Sciences}\ }\textbf {\bibinfo {volume} {109}},\ \bibinfo {pages} {16023}
  (\bibinfo {year} {2012})}\BibitemShut {NoStop}%
\bibitem [{\citenamefont {Kim}\ \emph {et~al.}(2020)\citenamefont {Kim},
  \citenamefont {Sau},\ and\ \citenamefont {Furst}}]{kim2020}%
  \BibitemOpen
  \bibfield  {author} {\bibinfo {author} {\bibfnamefont {H.}~\bibnamefont
  {Kim}}, \bibinfo {author} {\bibfnamefont {M.}~\bibnamefont {Sau}},\ and\
  \bibinfo {author} {\bibfnamefont {E.~M.}\ \bibnamefont {Furst}},\ }\bibfield
  {title} {\bibinfo {title} {An {{Expanded State Diagram}} for the {{Directed
  Self-Assembly}} of {{Colloidal Suspensions}} in {{Toggled Fields}}},\ }\href
  {https://doi.org/10.1021/acs.langmuir.0c01616} {\bibfield  {journal}
  {\bibinfo  {journal} {Langmuir}\ }\textbf {\bibinfo {volume} {36}},\ \bibinfo
  {pages} {9926} (\bibinfo {year} {2020})}\BibitemShut {NoStop}%
\bibitem [{\citenamefont {Sherman}\ and\ \citenamefont
  {Swan}(2016)}]{sherman2016}%
  \BibitemOpen
  \bibfield  {author} {\bibinfo {author} {\bibfnamefont {Z.~M.}\ \bibnamefont
  {Sherman}}\ and\ \bibinfo {author} {\bibfnamefont {J.~W.}\ \bibnamefont
  {Swan}},\ }\bibfield  {title} {\bibinfo {title} {Dynamic, {{Directed
  Self-Assembly}} of {{Nanoparticles}} {\emph{via}} {{Toggled Interactions}}},\
  }\href {https://doi.org/10.1021/acsnano.6b01050} {\bibfield  {journal}
  {\bibinfo  {journal} {ACS Nano}\ }\textbf {\bibinfo {volume} {10}},\ \bibinfo
  {pages} {5260} (\bibinfo {year} {2016})}\BibitemShut {NoStop}%
\bibitem [{\citenamefont {Sherman}\ and\ \citenamefont
  {Swan}(2019)}]{sherman2019c}%
  \BibitemOpen
  \bibfield  {author} {\bibinfo {author} {\bibfnamefont {Z.~M.}\ \bibnamefont
  {Sherman}}\ and\ \bibinfo {author} {\bibfnamefont {J.~W.}\ \bibnamefont
  {Swan}},\ }\bibfield  {title} {\bibinfo {title} {Transmutable {{Colloidal
  Crystals}} and {{Active Phase Separation}} {\emph{via}} {{Dynamic}},
  {{Directed Self-Assembly}} with {{Toggled External Fields}}},\ }\href
  {https://doi.org/10.1021/acsnano.8b08076} {\bibfield  {journal} {\bibinfo
  {journal} {ACS Nano}\ }\textbf {\bibinfo {volume} {13}},\ \bibinfo {pages}
  {764} (\bibinfo {year} {2019})}\BibitemShut {NoStop}%
\bibitem [{\citenamefont {Kim}\ \emph {et~al.}(2019)\citenamefont {Kim},
  \citenamefont {Bauer}, \citenamefont {Vasquez},\ and\ \citenamefont
  {Furst}}]{kim2019}%
  \BibitemOpen
  \bibfield  {author} {\bibinfo {author} {\bibfnamefont {H.}~\bibnamefont
  {Kim}}, \bibinfo {author} {\bibfnamefont {J.~L.}\ \bibnamefont {Bauer}},
  \bibinfo {author} {\bibfnamefont {P.~A.}\ \bibnamefont {Vasquez}},\ and\
  \bibinfo {author} {\bibfnamefont {E.~M.}\ \bibnamefont {Furst}},\ }\bibfield
  {title} {\bibinfo {title} {Structural coarsening of magnetic ellipsoid
  particle suspensions driven in toggled fields},\ }\href
  {https://doi.org/10.1088/1361-6463/ab062f} {\bibfield  {journal} {\bibinfo
  {journal} {Journal of Physics D: Applied Physics}\ }\textbf {\bibinfo
  {volume} {52}},\ \bibinfo {pages} {184002} (\bibinfo {year}
  {2019})}\BibitemShut {NoStop}%
\bibitem [{\citenamefont {Furst}(2025)}]{PSI-177}%
  \BibitemOpen
  \bibfield  {author} {\bibinfo {author} {\bibfnamefont {E.~M.}\ \bibnamefont
  {Furst}},\ }\href {https://doi.org/10.60555/jq1d-x679} {\bibinfo {title}
  {Investigating the structure of paramagnetic aggregates from colloidal
  emulsions -- 4 ({InSPACE}-4)}},\ \bibinfo {howpublished} {{NASA PSI} Data
  Repository} (\bibinfo {year} {2025}),\ \bibinfo {note}
  {\doi{10.60555/jq1d-x679}}\BibitemShut {NoStop}%
\bibitem [{\citenamefont {Whitmer}\ and\ \citenamefont
  {Luijten}(2011)}]{whitmer2011}%
  \BibitemOpen
  \bibfield  {author} {\bibinfo {author} {\bibfnamefont {J.~K.}\ \bibnamefont
  {Whitmer}}\ and\ \bibinfo {author} {\bibfnamefont {E.}~\bibnamefont
  {Luijten}},\ }\bibfield  {title} {\bibinfo {title} {Sedimentation of
  aggregating colloids},\ }\href {https://doi.org/10.1063/1.3525923} {\bibfield
   {journal} {\bibinfo  {journal} {The Journal of Chemical Physics}\ }\textbf
  {\bibinfo {volume} {134}},\ \bibinfo {pages} {034510} (\bibinfo {year}
  {2011})}\BibitemShut {NoStop}%
\bibitem [{\citenamefont {Vicsek}\ and\ \citenamefont
  {Family}(1984)}]{vicsek1984}%
  \BibitemOpen
  \bibfield  {author} {\bibinfo {author} {\bibfnamefont {T.}~\bibnamefont
  {Vicsek}}\ and\ \bibinfo {author} {\bibfnamefont {F.}~\bibnamefont
  {Family}},\ }\bibfield  {title} {\bibinfo {title} {Dynamic {{Scaling}} for
  {{Aggregation}} of {{Clusters}}},\ }\href
  {https://doi.org/10.1103/PhysRevLett.52.1669} {\bibfield  {journal} {\bibinfo
   {journal} {Physical Review Letters}\ }\textbf {\bibinfo {volume} {52}},\
  \bibinfo {pages} {1669} (\bibinfo {year} {1984})}\BibitemShut {NoStop}%
\bibitem [{\citenamefont {Conradt}\ and\ \citenamefont
  {Furst}(2025)}]{conradt2025}%
  \BibitemOpen
  \bibfield  {author} {\bibinfo {author} {\bibfnamefont {J.}~\bibnamefont
  {Conradt}}\ and\ \bibinfo {author} {\bibfnamefont {E.~M.}\ \bibnamefont
  {Furst}},\ }\bibfield  {title} {\bibinfo {title} {Quantitative {{Imaging}} of
  {{Colloidal Structures}}},\ }\href
  {https://doi.org/10.1021/acs.langmuir.4c05270} {\bibfield  {journal}
  {\bibinfo  {journal} {Langmuir}\ }\textbf {\bibinfo {volume} {41}},\ \bibinfo
  {pages} {8176} (\bibinfo {year} {2025})}\BibitemShut {NoStop}%
\bibitem [{\citenamefont {Edelsbrunner}\ and\ \citenamefont
  {M{\"u}cke}(1994)}]{edelsbrunner1994}%
  \BibitemOpen
  \bibfield  {author} {\bibinfo {author} {\bibfnamefont {H.}~\bibnamefont
  {Edelsbrunner}}\ and\ \bibinfo {author} {\bibfnamefont {E.~P.}\ \bibnamefont
  {M{\"u}cke}},\ }\bibfield  {title} {\bibinfo {title} {Three-dimensional alpha
  shapes},\ }\href {https://doi.org/10.1145/174462.156635} {\bibfield
  {journal} {\bibinfo  {journal} {ACM Transactions on Graphics}\ }\textbf
  {\bibinfo {volume} {13}},\ \bibinfo {pages} {43} (\bibinfo {year}
  {1994})}\BibitemShut {NoStop}%
\bibitem [{\citenamefont {Gardiner}\ \emph {et~al.}(2018)\citenamefont
  {Gardiner}, \citenamefont {Behnsen},\ and\ \citenamefont
  {Brassey}}]{gardiner2018}%
  \BibitemOpen
  \bibfield  {author} {\bibinfo {author} {\bibfnamefont {J.~D.}\ \bibnamefont
  {Gardiner}}, \bibinfo {author} {\bibfnamefont {J.}~\bibnamefont {Behnsen}},\
  and\ \bibinfo {author} {\bibfnamefont {C.~A.}\ \bibnamefont {Brassey}},\
  }\bibfield  {title} {\bibinfo {title} {Alpha shapes: Determining {{3D}} shape
  complexity across morphologically diverse structures},\ }\href
  {https://doi.org/10.1186/s12862-018-1305-z} {\bibfield  {journal} {\bibinfo
  {journal} {BMC Evolutionary Biology}\ }\textbf {\bibinfo {volume} {18}},\
  \bibinfo {pages} {184} (\bibinfo {year} {2018})}\BibitemShut {NoStop}%
\bibitem [{\citenamefont {Asaeedi}\ \emph {et~al.}(2017)\citenamefont
  {Asaeedi}, \citenamefont {Didehvar},\ and\ \citenamefont
  {Mohades}}]{asaeedi2017}%
  \BibitemOpen
  \bibfield  {author} {\bibinfo {author} {\bibfnamefont {S.}~\bibnamefont
  {Asaeedi}}, \bibinfo {author} {\bibfnamefont {F.}~\bibnamefont {Didehvar}},\
  and\ \bibinfo {author} {\bibfnamefont {A.}~\bibnamefont {Mohades}},\
  }\bibfield  {title} {\bibinfo {title} {{$\alpha$}-{{Concave}} hull, a
  generalization of convex hull},\ }\href
  {https://doi.org/10.1016/j.tcs.2017.08.014} {\bibfield  {journal} {\bibinfo
  {journal} {Theoretical Computer Science}\ }\textbf {\bibinfo {volume}
  {702}},\ \bibinfo {pages} {48} (\bibinfo {year} {2017})}\BibitemShut
  {NoStop}%
\bibitem [{\citenamefont {Borgoni}\ \emph {et~al.}(2021)\citenamefont
  {Borgoni}, \citenamefont {Galimberti},\ and\ \citenamefont
  {Zappa}}]{borgoni2021}%
  \BibitemOpen
  \bibfield  {author} {\bibinfo {author} {\bibfnamefont {R.}~\bibnamefont
  {Borgoni}}, \bibinfo {author} {\bibfnamefont {C.}~\bibnamefont
  {Galimberti}},\ and\ \bibinfo {author} {\bibfnamefont {D.}~\bibnamefont
  {Zappa}},\ }\bibfield  {title} {\bibinfo {title} {Identification of spatial
  defects in semiconductor manufacturing},\ }\href
  {https://doi.org/10.1002/asmb.2615} {\bibfield  {journal} {\bibinfo
  {journal} {Applied Stochastic Models in Business and Industry}\ }\textbf
  {\bibinfo {volume} {37}},\ \bibinfo {pages} {878} (\bibinfo {year}
  {2021})}\BibitemShut {NoStop}%
\bibitem [{\citenamefont {Blott}\ and\ \citenamefont {Pye}(2008)}]{blott2008}%
  \BibitemOpen
  \bibfield  {author} {\bibinfo {author} {\bibfnamefont {S.~J.}\ \bibnamefont
  {Blott}}\ and\ \bibinfo {author} {\bibfnamefont {K.}~\bibnamefont {Pye}},\
  }\bibfield  {title} {\bibinfo {title} {Particle shape: A review and new
  methods of characterization and classification},\ }\href
  {https://doi.org/10.1111/j.1365-3091.2007.00892.x} {\bibfield  {journal}
  {\bibinfo  {journal} {Sedimentology}\ }\textbf {\bibinfo {volume} {55}},\
  \bibinfo {pages} {31} (\bibinfo {year} {2008})}\BibitemShut {NoStop}%
\bibitem [{\citenamefont {Nye}(1984)}]{nye1984}%
  \BibitemOpen
  \bibfield  {author} {\bibinfo {author} {\bibfnamefont {J.~F.}\ \bibnamefont
  {Nye}},\ }\href@noop {} {\emph {\bibinfo {title} {Physical Properties of
  Crystals: Their Representation by Tensors and Matrices}}},\ \bibinfo
  {edition} {1st}\ ed.\ (\bibinfo  {publisher} {Clarendon Press ; Oxford
  University Press},\ \bibinfo {address} {Oxford [Oxfordshire] : New York},\
  \bibinfo {year} {1984})\BibitemShut {NoStop}%
\bibitem [{\citenamefont {Jackson}(2003)}]{jackson2003}%
  \BibitemOpen
  \bibfield  {author} {\bibinfo {author} {\bibfnamefont {J.~D.}\ \bibnamefont
  {Jackson}},\ }\bibfield  {title} {\bibinfo {title} {Electrodynamics,
  {{Classical}}},\ }in\ \href {https://doi.org/10.1002/3527600434.eap109}
  {\emph {\bibinfo {booktitle} {Digital {{Encyclopedia}} of {{Applied
  Physics}}}}},\ \bibinfo {editor} {edited by\ \bibinfo {editor} {\bibnamefont
  {{Wiley-VCH Verlag GmbH \& Co. KGaA}}}}\ (\bibinfo  {publisher} {Wiley},\
  \bibinfo {year} {2003})\ \bibinfo {edition} {1st}\ ed.\BibitemShut {Stop}%
\bibitem [{\citenamefont {Gonzalez}\ and\ \citenamefont
  {Woods}(2018)}]{gonzalez2018}%
  \BibitemOpen
  \bibfield  {author} {\bibinfo {author} {\bibfnamefont {R.~C.}\ \bibnamefont
  {Gonzalez}}\ and\ \bibinfo {author} {\bibfnamefont {R.~E.}\ \bibnamefont
  {Woods}},\ }\href@noop {} {\emph {\bibinfo {title} {Digital Image
  Processing}}}\ (\bibinfo  {publisher} {Pearson},\ \bibinfo {address} {New
  York, NY},\ \bibinfo {year} {2018})\BibitemShut {NoStop}%
\bibitem [{\citenamefont {Fitzgibbon}\ \emph {et~al.}(1999)\citenamefont
  {Fitzgibbon}, \citenamefont {Pilu},\ and\ \citenamefont
  {Fisher}}]{fitzgibbon1999}%
  \BibitemOpen
  \bibfield  {author} {\bibinfo {author} {\bibfnamefont {A.}~\bibnamefont
  {Fitzgibbon}}, \bibinfo {author} {\bibfnamefont {M.}~\bibnamefont {Pilu}},\
  and\ \bibinfo {author} {\bibfnamefont {R.}~\bibnamefont {Fisher}},\
  }\bibfield  {title} {\bibinfo {title} {Direct least square fitting of
  ellipses},\ }\href {https://doi.org/10.1109/34.765658} {\bibfield  {journal}
  {\bibinfo  {journal} {IEEE Transactions on Pattern Analysis and Machine
  Intelligence}\ }\textbf {\bibinfo {volume} {21}},\ \bibinfo {pages} {476}
  (\bibinfo {year} {1999})}\BibitemShut {NoStop}%
\bibitem [{\citenamefont {Ikotun}\ \emph {et~al.}(2025)\citenamefont {Ikotun},
  \citenamefont {Habyarimana},\ and\ \citenamefont {Ezugwu}}]{ikotun2025}%
  \BibitemOpen
  \bibfield  {author} {\bibinfo {author} {\bibfnamefont {A.~M.}\ \bibnamefont
  {Ikotun}}, \bibinfo {author} {\bibfnamefont {F.}~\bibnamefont
  {Habyarimana}},\ and\ \bibinfo {author} {\bibfnamefont {A.~E.}\ \bibnamefont
  {Ezugwu}},\ }\bibfield  {title} {\bibinfo {title} {Cluster validity indices
  for automatic clustering: {{A}} comprehensive review},\ }\href
  {https://doi.org/10.1016/j.heliyon.2025.e41953} {\bibfield  {journal}
  {\bibinfo  {journal} {Heliyon}\ }\textbf {\bibinfo {volume} {11}},\ \bibinfo
  {pages} {e41953} (\bibinfo {year} {2025})}\BibitemShut {NoStop}%
\bibitem [{\citenamefont {Xie}\ and\ \citenamefont {Beni}(1991)}]{xie1991}%
  \BibitemOpen
  \bibfield  {author} {\bibinfo {author} {\bibfnamefont {X.}~\bibnamefont
  {Xie}}\ and\ \bibinfo {author} {\bibfnamefont {G.}~\bibnamefont {Beni}},\
  }\bibfield  {title} {\bibinfo {title} {A validity measure for fuzzy
  clustering},\ }\href {https://doi.org/10.1109/34.85677} {\bibfield  {journal}
  {\bibinfo  {journal} {IEEE Transactions on Pattern Analysis and Machine
  Intelligence}\ }\textbf {\bibinfo {volume} {13}},\ \bibinfo {pages} {841}
  (\bibinfo {year} {1991})}\BibitemShut {NoStop}%
\bibitem [{\citenamefont {Promislow}\ and\ \citenamefont
  {Gast}(1997)}]{promislow1997c}%
  \BibitemOpen
  \bibfield  {author} {\bibinfo {author} {\bibfnamefont {J.~H.~E.}\
  \bibnamefont {Promislow}}\ and\ \bibinfo {author} {\bibfnamefont {A.~P.}\
  \bibnamefont {Gast}},\ }\bibfield  {title} {\bibinfo {title} {Low-energy
  suspension structure of a magnetorheological fluid},\ }\href
  {https://doi.org/10.1103/PhysRevE.56.642} {\bibfield  {journal} {\bibinfo
  {journal} {Physical Review E}\ }\textbf {\bibinfo {volume} {56}},\ \bibinfo
  {pages} {642} (\bibinfo {year} {1997})}\BibitemShut {NoStop}%
\bibitem [{\citenamefont {Semwal}\ \emph {et~al.}(2022)\citenamefont {Semwal},
  \citenamefont {{Clowe-Coish}}, \citenamefont {{Saika-Voivod}},\ and\
  \citenamefont {Yethiraj}}]{semwal2022}%
  \BibitemOpen
  \bibfield  {author} {\bibinfo {author} {\bibfnamefont {S.}~\bibnamefont
  {Semwal}}, \bibinfo {author} {\bibfnamefont {C.}~\bibnamefont
  {{Clowe-Coish}}}, \bibinfo {author} {\bibfnamefont {I.}~\bibnamefont
  {{Saika-Voivod}}},\ and\ \bibinfo {author} {\bibfnamefont {A.}~\bibnamefont
  {Yethiraj}},\ }\bibfield  {title} {\bibinfo {title} {Tunable {{Colloids}}
  with {{Dipolar}} and {{Depletion Interactions}}: {{Toward Field-Switchable
  Crystals}} and {{Gels}}},\ }\href
  {https://doi.org/10.1103/PhysRevX.12.041021} {\bibfield  {journal} {\bibinfo
  {journal} {Physical Review X}\ }\textbf {\bibinfo {volume} {12}},\ \bibinfo
  {pages} {041021} (\bibinfo {year} {2022})}\BibitemShut {NoStop}%
\bibitem [{\citenamefont {Rendos}\ \emph {et~al.}(2022)\citenamefont {Rendos},
  \citenamefont {Cao}, \citenamefont {Chern}, \citenamefont {Lauricella},
  \citenamefont {Succi}, \citenamefont {Werner}, \citenamefont {Dennis},\ and\
  \citenamefont {Brown}}]{rendos2022}%
  \BibitemOpen
  \bibfield  {author} {\bibinfo {author} {\bibfnamefont {A.}~\bibnamefont
  {Rendos}}, \bibinfo {author} {\bibfnamefont {W.}~\bibnamefont {Cao}},
  \bibinfo {author} {\bibfnamefont {M.}~\bibnamefont {Chern}}, \bibinfo
  {author} {\bibfnamefont {M.}~\bibnamefont {Lauricella}}, \bibinfo {author}
  {\bibfnamefont {S.}~\bibnamefont {Succi}}, \bibinfo {author} {\bibfnamefont
  {J.~G.}\ \bibnamefont {Werner}}, \bibinfo {author} {\bibfnamefont {A.~M.}\
  \bibnamefont {Dennis}},\ and\ \bibinfo {author} {\bibfnamefont {K.~A.}\
  \bibnamefont {Brown}},\ }\bibfield  {title} {\bibinfo {title} {Electric field
  induced macroscopic cellular phase of nanoparticles},\ }\href
  {https://doi.org/10.1039/D1SM01650D} {\bibfield  {journal} {\bibinfo
  {journal} {Soft Matter}\ }\textbf {\bibinfo {volume} {18}},\ \bibinfo {pages}
  {1991} (\bibinfo {year} {2022})}\BibitemShut {NoStop}%
\bibitem [{\citenamefont {Bauer}\ \emph {et~al.}(2015)\citenamefont {Bauer},
  \citenamefont {Liu}, \citenamefont {Kurian}, \citenamefont {Swan},\ and\
  \citenamefont {Furst}}]{bauer2015}%
  \BibitemOpen
  \bibfield  {author} {\bibinfo {author} {\bibfnamefont {J.~L.}\ \bibnamefont
  {Bauer}}, \bibinfo {author} {\bibfnamefont {Y.}~\bibnamefont {Liu}}, \bibinfo
  {author} {\bibfnamefont {M.~J.}\ \bibnamefont {Kurian}}, \bibinfo {author}
  {\bibfnamefont {J.~W.}\ \bibnamefont {Swan}},\ and\ \bibinfo {author}
  {\bibfnamefont {E.~M.}\ \bibnamefont {Furst}},\ }\bibfield  {title} {\bibinfo
  {title} {Coarsening mechanics of a colloidal suspension in toggled fields},\
  }\href {https://doi.org/10.1063/1.4927563} {\bibfield  {journal} {\bibinfo
  {journal} {The Journal of Chemical Physics}\ }\textbf {\bibinfo {volume}
  {143}},\ \bibinfo {pages} {074901} (\bibinfo {year} {2015})}\BibitemShut
  {NoStop}%
\bibitem [{\citenamefont {Bauer}\ \emph {et~al.}(2016)\citenamefont {Bauer},
  \citenamefont {Kurian}, \citenamefont {Stauffer},\ and\ \citenamefont
  {Furst}}]{bauer2016}%
  \BibitemOpen
  \bibfield  {author} {\bibinfo {author} {\bibfnamefont {J.~L.}\ \bibnamefont
  {Bauer}}, \bibinfo {author} {\bibfnamefont {M.~J.}\ \bibnamefont {Kurian}},
  \bibinfo {author} {\bibfnamefont {J.}~\bibnamefont {Stauffer}},\ and\
  \bibinfo {author} {\bibfnamefont {E.~M.}\ \bibnamefont {Furst}},\ }\bibfield
  {title} {\bibinfo {title} {Suppressing the {{Rayleigh}}--{{Plateau
  Instability}} in {{Field-Directed Colloidal Assembly}}},\ }\href
  {https://doi.org/10.1021/acs.langmuir.6b00771} {\bibfield  {journal}
  {\bibinfo  {journal} {Langmuir}\ }\textbf {\bibinfo {volume} {32}},\ \bibinfo
  {pages} {6618} (\bibinfo {year} {2016})}\BibitemShut {NoStop}%
\bibitem [{\citenamefont {Sherman}\ \emph {et~al.}(2018)\citenamefont
  {Sherman}, \citenamefont {Ghosh},\ and\ \citenamefont {Swan}}]{sherman2018}%
  \BibitemOpen
  \bibfield  {author} {\bibinfo {author} {\bibfnamefont {Z.~M.}\ \bibnamefont
  {Sherman}}, \bibinfo {author} {\bibfnamefont {D.}~\bibnamefont {Ghosh}},\
  and\ \bibinfo {author} {\bibfnamefont {J.~W.}\ \bibnamefont {Swan}},\
  }\bibfield  {title} {\bibinfo {title} {Field-{{Directed Self-Assembly}} of
  {{Mutually Polarizable Nanoparticles}}},\ }\href
  {https://doi.org/10.1021/acs.langmuir.8b01135} {\bibfield  {journal}
  {\bibinfo  {journal} {Langmuir}\ }\textbf {\bibinfo {volume} {34}},\ \bibinfo
  {pages} {7117} (\bibinfo {year} {2018})}\BibitemShut {NoStop}%
\bibitem [{\citenamefont {Risbud}\ and\ \citenamefont
  {Swan}(2015)}]{risbud2015}%
  \BibitemOpen
  \bibfield  {author} {\bibinfo {author} {\bibfnamefont {S.~R.}\ \bibnamefont
  {Risbud}}\ and\ \bibinfo {author} {\bibfnamefont {J.~W.}\ \bibnamefont
  {Swan}},\ }\bibfield  {title} {\bibinfo {title} {Dynamic self-assembly of
  colloids through periodic variation of inter-particle potentials},\ }\href
  {https://doi.org/10.1039/C5SM00185D} {\bibfield  {journal} {\bibinfo
  {journal} {Soft Matter}\ }\textbf {\bibinfo {volume} {11}},\ \bibinfo {pages}
  {3232} (\bibinfo {year} {2015})}\BibitemShut {NoStop}%
\bibitem [{\citenamefont {Guell}\ \emph {et~al.}(1988)\citenamefont {Guell},
  \citenamefont {Brenner}, \citenamefont {Frankel},\ and\ \citenamefont
  {Hartman}}]{guell1988}%
  \BibitemOpen
  \bibfield  {author} {\bibinfo {author} {\bibfnamefont {D.}~\bibnamefont
  {Guell}}, \bibinfo {author} {\bibfnamefont {H.}~\bibnamefont {Brenner}},
  \bibinfo {author} {\bibfnamefont {R.}~\bibnamefont {Frankel}},\ and\ \bibinfo
  {author} {\bibfnamefont {H.}~\bibnamefont {Hartman}},\ }\bibfield  {title}
  {\bibinfo {title} {Hydrodynamic forces and band formation in swimming
  magnetotactic bacteria},\ }\href
  {https://doi.org/10.1016/S0022-5193(88)80274-1} {\bibfield  {journal}
  {\bibinfo  {journal} {Journal of Theoretical Biology}\ }\textbf {\bibinfo
  {volume} {135}},\ \bibinfo {pages} {525} (\bibinfo {year}
  {1988})}\BibitemShut {NoStop}%
\bibitem [{\citenamefont {Kach}\ \emph {et~al.}(2024)\citenamefont {Kach},
  \citenamefont {Walker},\ and\ \citenamefont {Khair}}]{kach2024}%
  \BibitemOpen
  \bibfield  {author} {\bibinfo {author} {\bibfnamefont {J.~I.}\ \bibnamefont
  {Kach}}, \bibinfo {author} {\bibfnamefont {L.~M.}\ \bibnamefont {Walker}},\
  and\ \bibinfo {author} {\bibfnamefont {A.~S.}\ \bibnamefont {Khair}},\
  }\bibfield  {title} {\bibinfo {title} {A phenomenological model for chains
  and bands in dipolar suspensions},\ }\href
  {https://doi.org/10.1122/8.0000823} {\bibfield  {journal} {\bibinfo
  {journal} {Journal of Rheology}\ }\textbf {\bibinfo {volume} {68}},\ \bibinfo
  {pages} {581} (\bibinfo {year} {2024})}\BibitemShut {NoStop}%
\bibitem [{\citenamefont {Gast}\ and\ \citenamefont
  {Monovoukas}(1991)}]{gast1991}%
  \BibitemOpen
  \bibfield  {author} {\bibinfo {author} {\bibfnamefont {A.~P.}\ \bibnamefont
  {Gast}}\ and\ \bibinfo {author} {\bibfnamefont {Y.}~\bibnamefont
  {Monovoukas}},\ }\bibfield  {title} {\bibinfo {title} {A new growth
  instability in colloidal crystallization},\ }\href
  {https://doi.org/10.1038/351553a0} {\bibfield  {journal} {\bibinfo  {journal}
  {Nature}\ }\textbf {\bibinfo {volume} {351}},\ \bibinfo {pages} {553}
  (\bibinfo {year} {1991})}\BibitemShut {NoStop}%
\bibitem [{\citenamefont {Marr}\ and\ \citenamefont {Gast}(1994)}]{marr1994}%
  \BibitemOpen
  \bibfield  {author} {\bibinfo {author} {\bibfnamefont {D.~W.}\ \bibnamefont
  {Marr}}\ and\ \bibinfo {author} {\bibfnamefont {A.~P.}\ \bibnamefont
  {Gast}},\ }\bibfield  {title} {\bibinfo {title} {Interfacial {{Free Energy}}
  between {{Hard-Sphere Solids}} and {{Fluids}}},\ }\href
  {https://doi.org/10.1021/la00017a006} {\bibfield  {journal} {\bibinfo
  {journal} {Langmuir}\ }\textbf {\bibinfo {volume} {10}},\ \bibinfo {pages}
  {1348} (\bibinfo {year} {1994})}\BibitemShut {NoStop}%
\bibitem [{\citenamefont {Nichols}(1976)}]{nichols1976}%
  \BibitemOpen
  \bibfield  {author} {\bibinfo {author} {\bibfnamefont {F.~A.}\ \bibnamefont
  {Nichols}},\ }\bibfield  {title} {\bibinfo {title} {On the spheroidization of
  rod-shaped particles of finite length},\ }\href
  {https://doi.org/10.1007/BF02396641} {\bibfield  {journal} {\bibinfo
  {journal} {Journal of Materials Science}\ }\textbf {\bibinfo {volume} {11}},\
  \bibinfo {pages} {1077} (\bibinfo {year} {1976})}\BibitemShut {NoStop}%
\bibitem [{\citenamefont {Gorshkov}\ \emph {et~al.}(2019)\citenamefont
  {Gorshkov}, \citenamefont {Sareh}, \citenamefont {Tereshchuk},\ and\
  \citenamefont {Soleiman-Fallah}}]{gorshkov2019}%
  \BibitemOpen
  \bibfield  {author} {\bibinfo {author} {\bibfnamefont {V.~N.}\ \bibnamefont
  {Gorshkov}}, \bibinfo {author} {\bibfnamefont {P.}~\bibnamefont {Sareh}},
  \bibinfo {author} {\bibfnamefont {V.~V.}\ \bibnamefont {Tereshchuk}},\ and\
  \bibinfo {author} {\bibfnamefont {A.}~\bibnamefont {Soleiman-Fallah}},\
  }\bibfield  {title} {\bibinfo {title} {Dynamics of {{Anisotropic
  Break}}-{{Up}} in {{Nanowires}} of {{FCC Lattice Structure}}},\ }\href
  {https://doi.org/10.1002/adts.201900118} {\bibfield  {journal} {\bibinfo
  {journal} {Advanced Theory and Simulations}\ }\textbf {\bibinfo {volume}
  {2}},\ \bibinfo {pages} {1900118} (\bibinfo {year} {2019})}\BibitemShut
  {NoStop}%
\bibitem [{\citenamefont {Furst}\ and\ \citenamefont {Green}(2024)}]{PSI-74}%
  \BibitemOpen
  \bibfield  {author} {\bibinfo {author} {\bibfnamefont {E.~M.}\ \bibnamefont
  {Furst}}\ and\ \bibinfo {author} {\bibfnamefont {R.~D.}\ \bibnamefont
  {Green}},\ }\href {https://doi.org/10.60555/fxrc-a580} {\bibinfo {title}
  {Investigating the structures of paramagnetic aggregates from colloidal
  emulsions -- 2 ({InSPACE}-2)}},\ \bibinfo {howpublished} {{NASA PSI} Data
  Repository} (\bibinfo {year} {2024}),\ \bibinfo {note}
  {\doi{10.60555/fxrc-a580}}\BibitemShut {NoStop}%
\bibitem [{\citenamefont {Gast}\ and\ \citenamefont {Green}(2024)}]{PSI-76}%
  \BibitemOpen
  \bibfield  {author} {\bibinfo {author} {\bibfnamefont {A.~P.}\ \bibnamefont
  {Gast}}\ and\ \bibinfo {author} {\bibfnamefont {R.~D.}\ \bibnamefont
  {Green}},\ }\href {https://doi.org/10.60555/d3xq-e207} {\bibinfo {title}
  {Investigating the structures of paramagnetic aggregates from colloidal
  emulsions -- 1 ({InSPACE}-1)}},\ \bibinfo {howpublished} {{NASA PSI} Data
  Repository} (\bibinfo {year} {2024}),\ \bibinfo {note}
  {\doi{10.60555/d3xq-e207}}\BibitemShut {NoStop}%
\bibitem [{\citenamefont {Hynninen}\ and\ \citenamefont
  {Dijkstra}(2005)}]{hynninen2005}%
  \BibitemOpen
  \bibfield  {author} {\bibinfo {author} {\bibfnamefont {A.-P.}\ \bibnamefont
  {Hynninen}}\ and\ \bibinfo {author} {\bibfnamefont {M.}~\bibnamefont
  {Dijkstra}},\ }\bibfield  {title} {\bibinfo {title} {Phase {{Diagram}} of
  {{Dipolar Hard}} and {{Soft Spheres}}: {{Manipulation}} of {{Colloidal
  Crystal Structures}} by an {{External Field}}},\ }\href
  {https://doi.org/10.1103/PhysRevLett.94.138303} {\bibfield  {journal}
  {\bibinfo  {journal} {Physical Review Letters}\ }\textbf {\bibinfo {volume}
  {94}},\ \bibinfo {pages} {138303} (\bibinfo {year} {2005})}\BibitemShut
  {NoStop}%
\bibitem [{\citenamefont {Fermigier}\ and\ \citenamefont
  {Gast}(1992)}]{fermigier1992}%
  \BibitemOpen
  \bibfield  {author} {\bibinfo {author} {\bibfnamefont {M.}~\bibnamefont
  {Fermigier}}\ and\ \bibinfo {author} {\bibfnamefont {A.~P.}\ \bibnamefont
  {Gast}},\ }\bibfield  {title} {\bibinfo {title} {Structure evolution in a
  paramagnetic latex suspension},\ }\href
  {https://doi.org/10.1016/0021-9797(92)90165-I} {\bibfield  {journal}
  {\bibinfo  {journal} {Journal of Colloid and Interface Science}\ }\textbf
  {\bibinfo {volume} {154}},\ \bibinfo {pages} {522} (\bibinfo {year}
  {1992})}\BibitemShut {NoStop}%
\bibitem [{\citenamefont {Conradt}\ and\ \citenamefont
  {Furst}(2026)}]{conradt2026b}%
  \BibitemOpen
  \bibfield  {author} {\bibinfo {author} {\bibfnamefont {J.}~\bibnamefont
  {Conradt}}\ and\ \bibinfo {author} {\bibfnamefont {E.~M.}\ \bibnamefont
  {Furst}},\ }\bibfield  {title} {\bibinfo {title} {Magnetostatic energies in
  crystals of paramagnetic particles},\ }\href
  {https://doi.org/10.1103/pb3d-yg31} {\bibfield  {journal} {\bibinfo
  {journal} {Physical Review E}\ }\textbf {\bibinfo {volume} {113}},\ \bibinfo
  {pages} {015407} (\bibinfo {year} {2026})}\BibitemShut {NoStop}%
\bibitem [{\citenamefont {Conradt}\ \emph {et~al.}(2026)\citenamefont
  {Conradt}, \citenamefont {Sherman},\ and\ \citenamefont
  {Furst}}]{conradt2026}%
  \BibitemOpen
  \bibfield  {author} {\bibinfo {author} {\bibfnamefont {J.}~\bibnamefont
  {Conradt}}, \bibinfo {author} {\bibfnamefont {Z.~M.}\ \bibnamefont
  {Sherman}},\ and\ \bibinfo {author} {\bibfnamefont {E.~M.}\ \bibnamefont
  {Furst}},\ }\bibfield  {title} {\bibinfo {title} {Surface energies in
  crystals of mutually polarizing dipolar particles},\ }\href
  {https://doi.org/10.1063/5.0310699} {\bibfield  {journal} {\bibinfo
  {journal} {The Journal of Chemical Physics}\ }\textbf {\bibinfo {volume}
  {164}},\ \bibinfo {pages} {064707} (\bibinfo {year} {2026})}\BibitemShut
  {NoStop}%
\bibitem [{\citenamefont {Grasselli}\ \emph {et~al.}(1994)\citenamefont
  {Grasselli}, \citenamefont {Bossis},\ and\ \citenamefont
  {Lemaire}}]{grasselli1994}%
  \BibitemOpen
  \bibfield  {author} {\bibinfo {author} {\bibfnamefont {Y.}~\bibnamefont
  {Grasselli}}, \bibinfo {author} {\bibfnamefont {G.}~\bibnamefont {Bossis}},\
  and\ \bibinfo {author} {\bibfnamefont {E.}~\bibnamefont {Lemaire}},\
  }\bibfield  {title} {\bibinfo {title} {Structure induced in suspensions by a
  magnetic field},\ }\href {https://doi.org/10.1051/jp2:1994127} {\bibfield
  {journal} {\bibinfo  {journal} {Journal de Physique II}\ }\textbf {\bibinfo
  {volume} {4}},\ \bibinfo {pages} {253} (\bibinfo {year} {1994})}\BibitemShut
  {NoStop}%
\bibitem [{\citenamefont {Beleggia}\ \emph {et~al.}(2006)\citenamefont
  {Beleggia}, \citenamefont {De~Graef},\ and\ \citenamefont
  {Millev}}]{beleggia2006}%
  \BibitemOpen
  \bibfield  {author} {\bibinfo {author} {\bibfnamefont {M.}~\bibnamefont
  {Beleggia}}, \bibinfo {author} {\bibfnamefont {M.}~\bibnamefont {De~Graef}},\
  and\ \bibinfo {author} {\bibfnamefont {Y.}~\bibnamefont {Millev}},\
  }\bibfield  {title} {\bibinfo {title} {Demagnetization factors of the general
  ellipsoid: {{An}} alternative to the {{Maxwell}} approach},\ }\href
  {https://doi.org/10.1080/14786430600617161} {\bibfield  {journal} {\bibinfo
  {journal} {Philosophical Magazine}\ }\textbf {\bibinfo {volume} {86}},\
  \bibinfo {pages} {2451} (\bibinfo {year} {2006})}\BibitemShut {NoStop}%
\bibitem [{\citenamefont {Clercx}\ and\ \citenamefont
  {Bossis}(1993)}]{clercx1993}%
  \BibitemOpen
  \bibfield  {author} {\bibinfo {author} {\bibfnamefont {H.~J.~H.}\
  \bibnamefont {Clercx}}\ and\ \bibinfo {author} {\bibfnamefont
  {G.}~\bibnamefont {Bossis}},\ }\bibfield  {title} {\bibinfo {title}
  {Electrostatic interactions in slabs of polarizable particles},\ }\href@noop
  {} {\bibfield  {journal} {\bibinfo  {journal} {The Journal of Chemical
  Physics}\ }\textbf {\bibinfo {volume} {98}},\ \bibinfo {pages} {8284}
  (\bibinfo {year} {1993})}\BibitemShut {NoStop}%
\bibitem [{\citenamefont {Van~Rhee}\ \emph {et~al.}(2013)\citenamefont
  {Van~Rhee}, \citenamefont {Zijlstra}, \citenamefont {Verhagen}, \citenamefont
  {Aarts}, \citenamefont {Katsnelson}, \citenamefont {Maan}, \citenamefont
  {Orrit},\ and\ \citenamefont {Christianen}}]{vanrhee2013}%
  \BibitemOpen
  \bibfield  {author} {\bibinfo {author} {\bibfnamefont {P.~G.}\ \bibnamefont
  {Van~Rhee}}, \bibinfo {author} {\bibfnamefont {P.}~\bibnamefont {Zijlstra}},
  \bibinfo {author} {\bibfnamefont {T.~G.~A.}\ \bibnamefont {Verhagen}},
  \bibinfo {author} {\bibfnamefont {J.}~\bibnamefont {Aarts}}, \bibinfo
  {author} {\bibfnamefont {M.~I.}\ \bibnamefont {Katsnelson}}, \bibinfo
  {author} {\bibfnamefont {J.~C.}\ \bibnamefont {Maan}}, \bibinfo {author}
  {\bibfnamefont {M.}~\bibnamefont {Orrit}},\ and\ \bibinfo {author}
  {\bibfnamefont {P.~C.~M.}\ \bibnamefont {Christianen}},\ }\bibfield  {title}
  {\bibinfo {title} {Giant {{Magnetic Susceptibility}} of {{Gold Nanorods
  Detected}} by {{Magnetic Alignment}}},\ }\href
  {https://doi.org/10.1103/PhysRevLett.111.127202} {\bibfield  {journal}
  {\bibinfo  {journal} {Physical Review Letters}\ }\textbf {\bibinfo {volume}
  {111}},\ \bibinfo {pages} {127202} (\bibinfo {year} {2013})}\BibitemShut
  {NoStop}%
\bibitem [{\citenamefont {Martin}\ \emph {et~al.}(2008)\citenamefont {Martin},
  \citenamefont {Venturini},\ and\ \citenamefont {Huber}}]{martin2008}%
  \BibitemOpen
  \bibfield  {author} {\bibinfo {author} {\bibfnamefont {J.~E.}\ \bibnamefont
  {Martin}}, \bibinfo {author} {\bibfnamefont {E.}~\bibnamefont {Venturini}},\
  and\ \bibinfo {author} {\bibfnamefont {D.}~\bibnamefont {Huber}},\ }\bibfield
   {title} {\bibinfo {title} {Giant magnetic susceptibility enhancement in
  field-structured nanocomposites},\ }\href
  {https://doi.org/10.1016/j.jmmm.2008.04.111} {\bibfield  {journal} {\bibinfo
  {journal} {Journal of Magnetism and Magnetic Materials}\ }\textbf {\bibinfo
  {volume} {320}},\ \bibinfo {pages} {2221} (\bibinfo {year}
  {2008})}\BibitemShut {NoStop}%
\end{thebibliography}
%

\end{document}